%% file: volume-project.tex
\documentclass[final]{cdr}
\pdfoutput=1            
\graphicspath{ {figures/}{volume-project/figures/}{volume-project/generated/} }

\input{common/preamble}

\renewcommand\thedoctitle{\volintrotitle}
\def\titleextra{\includegraphics[width=0.95\textwidth]{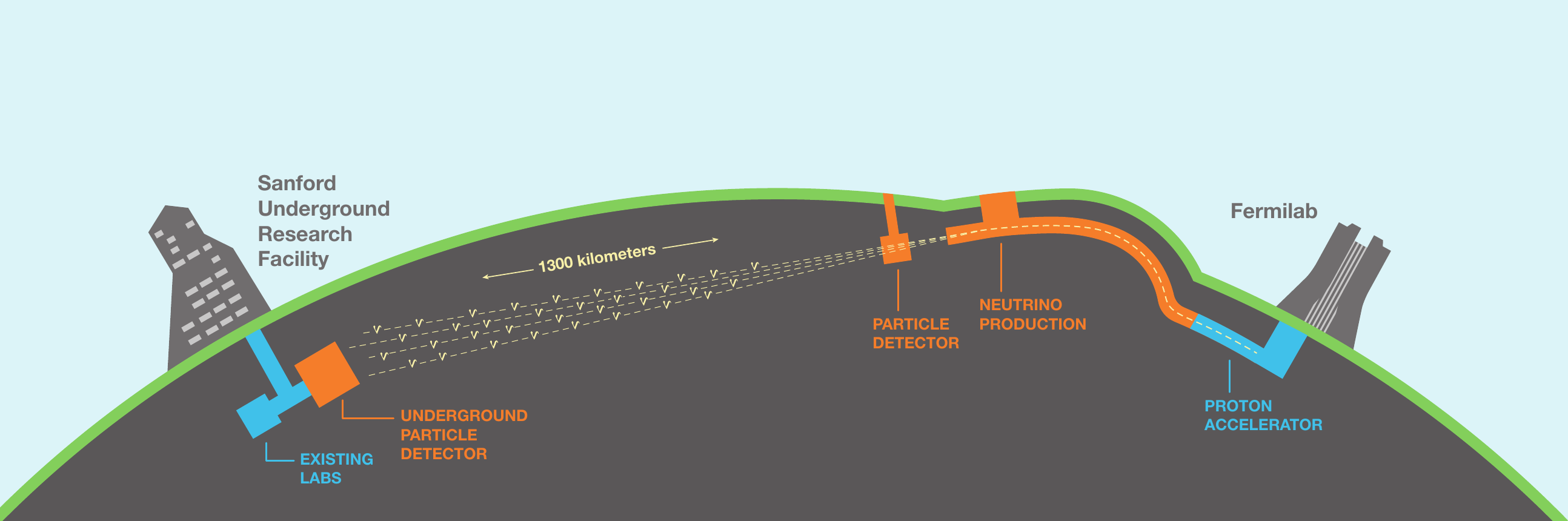}}
\begin{document}
\vspace{1in}

\input{common/init}

\input{common/acronyms-shared-vol1-2-4}
\input{common/acronyms-shared-vol1-4}
\input{common/acronyms-shared-vol1-2}
\input{common/acronyms-vol1}

\input{volume-project/chapter-overview}

\cleardoublepage

\input{volume-project/chapter-science}

\cleardoublepage

\input{volume-project/chapter-tech-designs}

\cleardoublepage

\input{volume-project/chapter-org-mgmt}

\cleardoublepage

\input{volume-project/chapter-summary}

\cleardoublepage

\input{common/final}
\end{document}

%% file: common/preamble.tex
\input{common/defs.tex}
\input{common/units.tex}

\input{generated/parameters.tex}
\hypersetup{
    pdftitle={\expshort CDR \thedocsubtitle},
    pdfauthor={\expshort Collaboration},
    final=true,
    colorlinks=false,
    linktocpage=true,
    linkbordercolor=blue,
    citebordercolor=green,
    urlbordercolor=magenta,
    filecolor=black,
    pdfpagemode=UseOutlines,
    pdfborderstyle={/S/U},  
}

%% file: common/defs.tex
\def\expshort{DUNE\xspace}
\def\explong{The Deep Underground Neutrino Experiment\xspace}

\def\thedocsubtitle{Long-Baseline Neutrino Facility (LBNF) and 
Deep Underground Neutrino Experiment (DUNE)}%
\def\cdrtitle{Conceptual Design Report}

\def\volintrotitle{Volume 1: The LBNF and DUNE Projects\xspace}

\def\volintro{Volume 1: \textit{The LBNF and DUNE Projects\xspace}}
\def\volphys{Volume 2: \textit{The Physics Program for DUNE at LBNF\xspace}}
\def\vollbnf{Volume 3: \textit{The Long-Baseline Neutrino Facility for DUNE\xspace}}
\def\voldune{Volume 4: \textit{The DUNE Detectors at LBNF\xspace}}

\newcommand{\sinstt}[1]{$\sin^22\theta_{#1}$\xspace}
\newcommand{\sinst}[1]{$\sin^2\theta_{#1}$\xspace}
\newcommand{\deltacp}{$\delta_{\rm CP}$\xspace}

\newcommand{\numu}{$\nu_\mu$\xspace}
\newcommand{\nue}{$\nu_e$\xspace}

\newcommand{\anumu}{$\bar\nu_\mu$\xspace}
\newcommand{\anue}{$\bar\nu_e$\xspace}

\newcommand{\mdeltacp}{\delta_{\rm CP}}

\def\Ar39{$^{39}$Ar}

%% file: common/units.tex
\DeclareSIUnit \kton {\kilo\tonne}
\DeclareSIUnit \kt {\kilo\tonne}
\DeclareSIUnit \Mt {\mega\tonne}
\DeclareSIUnit \eV {\electronvolt}
\DeclareSIUnit \keV {\kilo\electronvolt}
\DeclareSIUnit \MeV {\mega\electronvolt}
\DeclareSIUnit \GeV {\giga\electronvolt}
\DeclareSIUnit \km {\kilo\meter}
\DeclareSIUnit \kW {\kilo\watt}
\DeclareSIUnit \MW {\mega\watt}
\DeclareSIUnit \MHz {\mega\hertz}
\DeclareSIUnit \mrad {\milli\radian}
\DeclareSIUnit \year {year}
\DeclareSIUnit \POT {POT}
\DeclareSIUnit \sig {$\sigma$}
\DeclareSIUnit\parsec{pc}
\DeclareSIUnit\lightyear{ly}
\DeclareSIUnit\foot{ft}
\DeclareSIUnit\ft{ft}
\def\ktyr{\si[inter-unit-product=\ensuremath{{}\cdot{}}]{\kt\year}\xspace}

\def\ktMWyr{\si[inter-unit-product=\ensuremath{{}\cdot{}}]{\kt\MW\year}\xspace}

\newcommand{\SIadj}[2]{\SI[number-unit-product = -]{#1}{#2}}
\newcommand{\ktadj}[1]{\SIadj{#1}{\kt}}
\newcommand{\kmadj}[1]{\SIadj{#1}{\km}}

\newcommand{\GeVadj}[1]{\SIadj{#1}{\GeV}}
\newcommand{\MWadj}[1]{\SIadj{#1}{\MW}}



%% file: common/init.tex

\pagestyle{titlepage}

\begin{center}
   {\Huge  \thedocsubtitle}  

  \vspace{5mm}

  {\Huge  \cdrtitle}  

  \vspace{10mm}

 {\LARGE \thedoctitle}

  \vspace{15mm}

\titleextra

  \vspace{10mm}
  \today
    \vspace{15mm}
    
\end{center}

\cleardoublepage



# The DUNE Collaboration

R. Acciarri[43], M. A. Acero[10], M. Adamowski[43], C. Adams[144], P. Adamson[43], S. Adhikari[77], Z. Ahmad[135], C. H. Albright[43], T. Alion[120], E. Amador[131], J. Anderson[8], K. Anderson[43], C. Andreopoulos[76], M. Andrews[43], R. Andrews[43], I. Anghel[62], J. d. Anjos[99], A. Ankowski[131], M. Antonello[72], A. Aranda Fernandez[31], A. Ariga[13], T. Ariga[13], D. Aristizabal[75], E. Arrieta-Diaz[87], K. Aryal[131], J. Asaadi[129], D. Asner[103], M. S. Athar[6], M. Auger[13], A. Aurisano[29], V. Aushev[68], D. Autiero[63], M. Avila[131], J. J. Back[139], X. Bai[122], B. Baibussinov[106], M. Baird[61], A. B. Balantekin[142], B. Baller[43], P. Ballett[41], B. Bambah[52], M. Bansal[107], S. Bansal[107], G. J. Barker[139], W. A. Barletta[83], G. Barr[102], N. Barros[109], L. Bartoszek[78], A. Bashyal[131], M. Bass[102], F. Bay[133], J. Beacom[100], B. R. Behera[107], G. Bellettini[110], V. Bellini[26], O. Beltramello[16], P. A. Benetti[108], A. Bercellie[114], M. Bergevin[19], E. Berman[43], H. Berns[20], R. Bernstein[43], S. Bertolucci[16], B. Bhandari[49], V. Bhatnagar[107], B. Bhuyan[60], J. Bian[90], K. Biery[43], M. Bishai[15], T. Blackburn[128], A. Blake[73], F. d. M. Blaszczyk[14], E. Blaufuss[82], B. Bleakley[124], E. Blucher[28], V. Bocean[43], F. Boffelli[108], J. Boissevain[78], S. Bolognesi[117], T. Bolton[65], M. Bonesini[89], T. Boone[33], C. Booth[118], S. Bordoni[53], M. Borysova[68], B. Bourguille[53], S. B. Boyd[139], D. Brailsford[73], A. Brandt[131], J. Bremer[16], S. Brice[43], C. Bromberg[87], G. Brooijmans[34], G. Brown[131], R. Brown[15], G. Brunetti[43], X. Bu[43], N. Buchanan[33], H. Budd[114], B. Bugg[130], P. Calafiura[74], E. Calligarich[108], E. Calvo[17], L. Camilleri[34], M. Campanelli[77], C. Cantini[147], B. Carls[43], R. Carr[34], M. Cascella[77], C. Castromonte[44], E. Catano Mur[62], F. Cavanna[43], S. Centro[106], A. Cervera Villanueva[136], V. B. Chandratre[12], A. Chatterjee[131], S. Chattopadhyay[43], S. Chattopadhyay[135], L. Chaussard[63], S. Chembra[52], H. Chen[15], K. Chen[15], M. Chen[21], D. Cherdack[33], C. Chi[34], S. Childress[43], S. Choubey[47], B. C. Choudhary[38], G. Christodoulou[76], C. Christofferson[122], E. Church[103], D. Cianci[81], D. Cline[21], T. Coan[125], A. Cocco[93], J. Coelho[134], P. Cole[58], G. Collin[83], J. M. Conrad[83], M. Convery[115], R. Corey[122], L. Corwin[122], J. Cranshaw[8], P. Crivelli[147], D. Cronin-Hennessy[90], A. Curioni[89], J. Cushing[28], D. L. Adams[15], D. Dale[58], S. R. Das[60], T. Davenne[116], G. S. Davies[61], J. Davies[128], J. Dawson[2], K. De[131], A. de Gouvea[97], J. K. de Jong[102], P. de Jong[92], P. De Lurgio[8], M. Decowski[92], A. Delbart[117], C. Densham[116], R. Dharmapalan[8], N. Dhingra[107], S. Di Luise[147], M. Diamantopoulou[9], J. S. Diaz[61], G. Diaz Bautista[114], M. Diwan[15], Z. Djurcic[8], J. Dolph[15], G. Drake[8], D. Duchesneau[7], M. Duvernois[142], H. Duyang[120], D. A. Dwyer[74], S. Dye[48], S. Dytman[111], B. Eberly[115], R. Edgecock[50], D. Edmunds[87], S. Elliott[78], M. Elnimr[21], S. Emery[117], E. Enoffness[105], S. Eno[82], A. Ereditato[13], C. O. Escobar[43], J. Evans[81], A. Falcone[108], L. Falk[128], A. Farbin[131], C. Farnese[106], Y. Farzan[57], A. Fava[106], L. Favilli[18], J. Felde[82], J. Felix[46], S. Fernandes[4], L. Fields[43], A. Finch[73], M. Fitton[116], B. Fleming[144], T. Forest[58], J. Fowler[40], W. Fox[61], J. Fried[15], A. Friedland[78], S. Fuess[43], B. Fujikawa[74], A. Gago[105], H. Gallagher[134], S. Galymov[63], T. Gamble[118], R. Gandhi[47], D. Garcia-Gamez[81], S. Gardiner[20], G. Garvey[78], V. M. Gehman[74], A. Gendotti[147], G. d. Geronimo[15], C. Ghag[77], P. Ghoshal[47], D. Gibin[106], I. Gil-Botella[17], R. Gill[15], D. Girardelli[25], A. Giri[55], S. Glavin[109], D. Goelli[13], S. Golapinni[65], M. Gold[96], R. A. Gomes[44], J. J. Gomez Cadenas[136], M. C. Goodman[8], D. Gorbunov[56], S. Goswami[47], N. Graf[111], N. Graf[115], M. Graham[115], E. Gramelini[144], R. Gran[91], C. Grant[20], N. Grant[139], V. Greco[26], H. Greenlee[43], L. Greenler[142], C. Greenley[79], M. Groh[29], S. Grullon[109], T. Grundy[73], K. Grzelak[138], E. Guardincerri[78], V. Guarino[8], E. Guarnaccia[137], G. P. Guedes[42], R. Guenette[102], A. Guglielmi[106], A. T. Habig[91], R. W. Hackenburg[15], A. Hackenburg[144], H. Hadavand[131], R. Haenni[13], A. Hahn[43], M. D. Haigh[139], T. Haines[78], T. Hamernik[43], T. Handler[130], S. Hans[15], D. Harris[43], J. Hartnell[128], T. Hasegawa[67], R. Hatcher[43], A. Hatzikoutelis[130], S. Hays[43], E. Hazen[14], M. Headley[123], A. Heavey[43], K. Heeger[144], J. Heise[123], K. Hennessy[76], J. Hewes[81], A. Higuera[49], T. Hill[58], A. Himmel[43], M. Hogan[33], P. Holanda[25], A. Holin[77], W. Honey[73], S. Horikawa[147], G. Horton-Smith[65], B. Howard[61], J. Howell[43], P. Hurh[43], J. Huston[87], J. Hylen[43], R. Imlay[79], J. Insler[79], G. Introzzi[108], D. Ioanisyan[145], A. Ioannisian[145], K. Iwamoto[114], A. Izmaylov[136], C. Jackson[131], D. E. Jaffe[15], C. James[43], E. James[43], F. Jediny[36], C. Jen[137], A. Jhingan[107], S. Jiménez[17], J. H. Jo[17], M. Johnson[43], R. Johnson[29], J. Johnstone[43], B. J. Jones[83], J. Joshi[15], H. Jostlein[43], C. K. Jung[127], T. Junk[43], A. Kaboth[59], R. Kadel[74], T. Kafka[134], L. Kalousis[137], Y. Kamyshkov[130], G. Karagiorgi[81], D. Karasavvas[9], Y. Karyotakis[71], A. Kaur[107], P. Kaur[107], B. Kayser[43], N. Kazaryan[145], E. Kearns[14], P. Keener[109], S. Kemboi[131], E. Kemp[25], S. H. Kettell[15], M. Khabibullin[56], M. Khandaker[58], A. Khotjantsev[56], B. Kirby[15], M. Kirby[43], J. Klein[109], T. Kobilarcik[43], S. Kohn[19], G. Koizumi[43], A. Kopylov[56], M. Kordosky[141], L. Kormos[73], U. Kose[16], V. A. Kostelecky[61], M. Kramer[19], I. Kreslo[13], D. Kriske[90], W. Kropp[21], Y. Kudenko[56], V. A. Kudryavtsev[118], S. Kulagin[56], A. Kumar[107], G.K. Kumar[69], J. Kumar[107], T. Kutter[79], A. Laminack[4], K. Lande[109], C. Lane[39], K. Lang[132], F. Lanni[15], J. Learned[48], P. Lebrun[43], D. Lee[78], H. Lee[114], K. Lee[21], W. M. Lee[43], M. A. Leigui de Oliveira[1], Q. Li[43], S. Li[15], S. Li[100], X. Li[127], Y. Li[15], Z. Li[40], J. Libo[120], C. S. Lin[74], S. Lin[33], J. Ling[15], J. Link[137], Z. Liptak[32], D. Lissauer[15], L. Littenberg[15], B. Littlejohn[54], Q. Liu[78], T. Liu[125], S. Lockwitz[43], N. Lockyer[43], T. Loew[74], M. Lokajicek[3], K. Long[59], M. D. L. Lopes[43], J. P. Lopez[32], J. Losecco[98], W. Louis[78],

J. Lowery[61], M. Luethi[13], K. B. Luk[19], B. Lundberg[43], T. Lundin[43], X. Luo[144], T. Lux[53], J. Lykken[43],
A. A. Machado[72], J. R. Macier[43], S. Magill[8], G. Mahler[15], K. Mahn[87], M. Malek[59], S. Malhotra[12], D. Malon[8],
F. Mammoliti[26], S. Mancina[8], S. K. Mandal[38], S. Mandodi[52], S. L. Manly[114], A. Mann[134], A. Marchionni[43],
W. Marciano[15], C. Mariani[137], J. Maricic[48], A. Marino[32], M. Marshak[90], C. Marshall[114], J. Marshall[24],
J. Marteau[63], J. Martin-Albo[102], D. Martinez[54], S. Matsuno[48], J. Matthews[79], C. Mauger[78], K. Mavrokoridis[76],
D. Mayilyan[145], E. Mazzucato[117], N. McCauley[76], E. McCluskey[43], N. McConkey[118], K. McDonald[112],
K. S. McFarland[114], A. M. McGowan[114], C. McGrew[127], R. McKeown[141], D. McNulty[58], R. McTaggart[124],
A. Mefodiev[56], M. Mehrian[33], P. Mehta[95], D. Mei[121], O. Mena[136], S. Menary[146], H. Mendez[85], A. Menegolli[108],
G. Meng[106], Y. Meng[21], H. Merritt[59], D. Mertins[4], M. Nessi[16], M. Mewes[61], W. Metcalf[79], M. Mewes[61], H. Meyer[140],
T. Miao[43], R. Milincic[48], W. Miller[90], G. Mills[78], O. Mineev[56], O. Miranda[30], C. S. Mishra[43], S. R. Mishra[120],
B. Mitrica[51], D. Mladenov[16], I. Mocioiu[104], R. Mohanta[52], N. Mokhov[43], C. Montanari[16], D. Montanari[43],
J. Moon[83], M. Mooney[15], C. Moore[43], J. Morfin[43], B. Morgan[139], C. Morris[49], W. Morse[15], Z. Moss[83], C. Mossey[43],
C. A. Moura[1], J. Mousseau[86], L. Mualem[23], M. Muether[140], S. Mufson[61], S. Murphy[147], J. Musser[61], R. Musser[131],
Y. Nakajima[74], D. Naples[111], J. Navarro[10], D. Navas[17], J. Nelson[141], M. Nessi[16], M. Newcomer[109], Y. Ng[131],
R. Nichol[77], T. C. Nicholls[116], K. Nikolics[147], E. Niner[61], B. Norris[43], F. Noto[16], P. Novakova[5], P. Novella[136],
J. Nowak[73], M. S. Nunes[25], H. O'Keeffe[73], R. Oldeman[24], R. Oliveira[25], T. Olson[134], V. Onishchuk[68], J. Osta[43],
T. Ovsjannikova[56], B. Page[87], S. Pakvasa[48], S. Pal[118], O. Palamara[43], A. Palazzo[84], J. Paley[43], C. Palomares[17],
E. Pantic[20], V. Paolone[111], V. Papadimitriou[43], J. Park[137], S. Parke[43], Z. Parsa[15], S. Pascoli[41], R. Patterson[23],
S. Patton[74], T. Patzak[2], B. Paulos[142], L. Paulucci[1], Z. Pavlovic[43], G. Pawloski[90], S. Peeters[128], E. Pennacchio[63],
A. Perch[77], G. N. Perdue[43], L. Periale[147], J. D. Perkin[118], H. Pessard[71], G. Petrillo[43], R. Petti[120], A. Petukhov[122],
F. Pietropaolo[106], R. Plunkett[43], S. Pordes[43], M. Potekhin[15], R. Potenza[26], B. Potukuchi[64], N. Poudyal[121],
O. Prokofiev[43], N. Pruthi[107], P. Przewlocki[94], D. Pushka[43], X. Qian[15], J. L. Raaf[43], R. Raboanary[7], V. Radeka[15],
A. Radovic[141], G. Raffelt[84], I. Rakhno[43], H. T. Rakotondramanana[7], L. Rakotondravohitra[7], Y. A. Ramachers[139],
R. Rameika[43], J. Ramsey[78], A. Rappoldi[108], G. Raselli[108], P. Ratoff[73], B. Rebel[43], C. Regenfus[147],
J. Reichenbacher[122], D. Reitzner[43], A. Remoto[71], A. Renshaw[49], S. Rescia[15], M. Richardson[118], K. Rielage[78],
K. Riesselmann[43], M. Robinson[118], L. Rochester[115], O. B. Rodrigues[25], P. Rodrigues[114], B. Roe[86], M. Rosen[48],
R. M. Roser[43], M. Ross-Lonergan[41], M. Rossella[108], A. Rubbia[147], C. Rubbia[45], R. Rucinski[43],
C. Rudolph von Rohr[13], B. Russell[144], J. Ruterbories[114], R. Saakyan[77], N. Sahu[55], P. Sala[88], N. Samios[15],
F. Sanchez[53], M. Sanchez[62], B. Sands[112], S. Santana[85], R. Santorelli[17], G. Santucci[127], N. Saoulidou[9],
A. Scaramelli[88], H. Schellman[101], P. Schlabach[43], R. Schmitt[43], D. Schmitz[28], J. Schneps[134], K. Scholberg[40],
A. Schukraft[43], J. Schwehr[33], E. Segreto[72], S. Seibert[109], J. A. Sepulveda-Quiroz[62], F. Sergiampietri[147],
L. Sexton-Kennedy[43], D. Sgalaberna[147], M. Shaevitz[34], J. Shahi[107], S. Shahsavarani[131], P. Shanahan[43],
S. U. Shankar[47], R. Sharma[15], R. K. Sharma[113], T. Shaw[43], R. Shrock[127], I. Shyrma[68], N. Simos[15], G. Sinev[40],
I. Singh[107], J. Singh[107], J. Singh[80], V. Singh[11], G. Sinnis[78], W. Sippach[34], D. Smargianaki[16], M. Smy[21],
E. Snider[43], P. Snopok[54], J. Sobczyk[143], H. Sobel[21], M. Soderberg[112], N. Solomey[140], W. Sondheim[78], M. Sorel[136],
A. Sousa[29], K. Soustruznik[27], J. Spitz[86], N. J. Spooner[118], M. Stancari[43], I. Stancu[4], D. Stefan[16], H. M. Steiner[74],
J. Stewart[15], J. Stock[122], S. Stoica[51], J. Stone[14], J. Strait[43], M. Strait[28], T. Strauss[43], S. Striganov[43], R. Sulej[94],
G. Sullivan[82], Y. Sun[48], L. Suter[8], C. M. Sutera[26], R. Svoboda[20], B. Szczerbinska[37], A. Szelc[81],
S. Söldner-Rembold[81], R. Talaga[8], M. Tamsett[128], S. Tariq[43], E. Tatar[58], R. Tayloe[61], C. Taylor[78], D. Taylor[123],
K. Terao[34], M. Thiesse[118], J. Thomas[77], L. F. Thompson[118], M. Thomson[24], C. Thorn[15], M. Thorpe[116], X. Tian[120],
D. Tiedt[122], S. C. Timm[43], A. Tonazzo[2], T. Tope[43], A. Topkar[12], F. R. Torres[25], M. Torti[108], M. Tortola[136],
F. Tortorici[26], M. Toups[43], C. Touramanis[76], M. Tripathi[20], I. Tropin[43], Y. Tsai[115], K. V. Tsang[74], R. Tsenov[119],
S. Tufanli[144], C. Tull[74], J. Turner[41], M. Tzanov[79], E. Tziaferi[9], Y. Uchida[59], J. Urheim[61], T. Usher[115], M. Vagins[66],
P. Vahle[141], G. A. Valdiviesso[5], L. Valerio[43], Z. Vallari[127], J. Valle[136], R. Van Berg[109], R. Van de Water[78],
P. Van Gemmeren[8], F. Varanini[106], G. Varner[48], G. Vasseur[117], K. Vaziri[43], G. Velev[43], S. Ventura[106],
A. Verdugo[17], T. Viant[147], T. V. Vieira[25], C. Vignoli[72], C. Vilela[127], B. Viren[15], T. Vrba[36], T. Wachala[70],
D. Wahl[142], M. Wallbank[118], N. Walsh[20], B. Wang[125], H. Wang[22], L. Wang[121], T. Wang[41], T. K. Warburton[118],
D. Warner[33], M. Wascko[59], D. Waters[77], T. B. Watson[131], A. Weber[116], M. Weber[13], W. Wei[121], A. Weinstein[62],
D. Wells[122], D. Wenman[142], M. Wetstein[62], A. White[131], L. Whitehead[49], D. Whittington[61], M. Wilking[127],
J. Willhite[43], P. Wilson[43], R. J. Wilson[33], L. Winslow[83], P. Wittich[35], S. Wojcicki[126], H. H. Wong[19], K. Wood[120],
E. Worcester[15], M. Worcester[15], S. Wu[147], T. Xin[62], C. Yanagisawa[127], S. Yang[29], T. Yang[43], K. Yarritu[78], J. Ye[125],
M. Yeh[15], N. Yershov[56], K. Yonehara[43], B. Yu[15], J. Yu[131], J. Zalesak[3], A. Zalewska[70], B. Zamorano[128], L. Zang[118],
A. Zani[16], G. Zavala[46], G. Zeller[43], C. Zhang[15], C. Zhang[121], E. D. Zimmerman[32], M. Zito[117], and R. Zwaska[43]

[1] *ABC Federal University, Santo André –SP, 09210-580, Brazil*
[2] *APC-Paris, Batiment Condorcet; 10, rue Alice Domon et L'eonie Duquet; F-75205 Paris CEDEX 13, France*
[3] *Institute of Physics ASCR, v. v. i. , Na Slovance 2; 182 21 Praha 8, Czech Republic*
[4] *University of Alabama (Tuscaloosa), Tuscaloosa, AL 35487-0324, USA*
[5] *University Federal de Alfenas em Poços de Caldas, 11999, CEP 37115-900 Poços de Caldas-MG, Brazil*

[6] *Aligarh Muslim University, Department of Physics Aligarh-202002, India*

[7] *Antananarivo, Présidence de l'Université d'Antananarivo: BP 566, Antananarivo 101, Madagascar*

[8] *Argonne National Lab., Argonne, IL 60439, USA*

[9] *University of Athens, University Campus, Zografou GR 157 84 Greece, Greece*

[10] *Universidad del Atlantico, Km 7 antigua vi a Puerto Colombia, Barranquilla, Colombia*

[11] *Banaras Hindu University, Dept. of Physics, Varanasi UP 221005, India*

[12] *Bhabha Atmoic Research Center, Trombay, Mumbai, India*

[13] *University of Bern, Lab for High Energy Physics; Sidlerstrasse 5; CH-3012 Bern, Switzerland*

[14] *Boston University, Boston, MA 02215, USA*

[15] *Brookhaven National Lab., Upton, NY 11973-5000, USA*

[16] *CERN, European Organization for Nuclear Research European Laboratory for Particle Physics; 1211 Gen'eve 23, Switzerland*

[17] *CIEMAT, División de Física de Partículas; Avenida Complutense 40; E-28040 Madrid, Spain*

[18] *CNR Pisa, 1, 56124 Pisa PI, Italy*

[19] *University of California (Berkeley), Berkeley, CA 94720-7300, USA*

[20] *University of California (Davis), Davis, CA 95616, USA*

[21] *University of California (Irvine), Irvine, CA 92697-4575, USA*

[22] *University of California (Los Angeles), Los Angeles, CA 90095-1547, USA*

[23] *California Inst. of Tech., MC 356-48; Pasadena, CA 91125, USA*

[24] *University of Cambridge, JJ Thomson Avenue, Cambridge CB3 0HE, UK*

[25] *University de Campinas, Av. Sérgio Buarque de Holanda, 777 CEP 13083-859 Campinas-SP, Brazil*

[26] *University di Catania, Via Santa Sofia, I-95123 Catania, Italy*

[27] *Institute of Particle and Nuclear Physics of the Faculty of Mathematics and Physics of the Charles University in Prague, V Holešovičkách 747/2; 180 00 Praha 8-Libeň, Czech Republic*

[28] *University of Chicago, Chicago, IL 60637-1434, USA*

[29] *University of Cincinnati, Cincinnati, OH 45221-0011, USA*

[30] *Cinvestav, 07360 Ciudad de México, D.F., Mexico*

[31] *Universidad de Colima, Facultad de Ciencias Bernal Diaz del Castillo 340 Colonia Villa San Sebastian Colima, Colima, Mexico*

[32] *University of Colorado, Boulder, CO 80309, USA*

[33] *Colorado State University, Fort Collins, CO 80523, USA*

[34] *Columbia University, New York, NY 10027, USA*

[35] *Cornell University, Laboratory for Elementary-Particle Physics; Newman Lab; Ithaca, NY 14853–5001, USA*

[36] *Czech Technical University in Prague, Brehova 7; 115 19 Praha 1, Czech Republic*

[37] *Dakota State University, Madison, SD 57042, USA*

[38] *University of Delhi, Department of Physics and Astrophysics, Delhi 110007, India*

[39] *Drexel University, Philadelphia, PA 19104, USA*

[40] *Duke University, Durham, NC 27706, USA*

[41] *University of Durham, Institute for Particle Physics Phenomenology; Dept. of Physics, Ogden Centre for Fund. Physics; South Road; Durham DH1 3LE, UK*

[42] *University Estadual de Feira de Santana, S/N;44036-900, Feira de Santana-BA, Brazil*

[43] *Fermi National Accelerator Lab, Batavia, IL 60510-0500, USA*

[44] *University Federal de Goias, Goiania, GO, Brazil*

[45] *Gran Sasso Science Institute, 7, L'Aquila, Italy*

[46] *Universidad de Guanajuato, Gto., C.P.37000, Mexico*

[47] *Harish-Chandra Research Institute, Jhunsi, Allahabad 211 019, India*

[48] *University of Hawaii, Honolulu, HI 96822-2219, USA*

[49] *University of Houston, Houston, TX 77204, USA*

[50] *Huddersfield, Huddersfield, West Yorkshire HD1 3DH, UK*

[51] *Horia Hulubei National Institute of Physiscs and Nuclear Engineering, Horia Hulubei National Institute of Physiscs and Nuclear Engineering, Bucharest-Magurele, Romania*

[52] *University of Hyderabad, Gachibowli, Hyderabad - 500 046, India*

[53] *Institut de Física d'Altes Energies (IFAE); Campus Universitat Autonoma de Barcelona , E-08193 Cerdanyola del Valles (Barcelona), Spain*

[54] *Illinois Institute of Technology , Room 182LS; Chicago, IL 60616, USA*

[55] *IIT Hyderabad, Kandi, Hyderabad - 502205, India*

[56] *Institute for Nuclear Research of the Russian Academy of Sciences (INR RAS), 60th October Anniversary Prosp. 7a Moscow, Russia 117312, Russia*

[57] *Institute for Research in Fundamental Sciences (IPM), Tehran, Iran. Postal code: 19538-33511 PO Box:*

3

*19395-5531, Iran*

[58]*Idaho State University, Department of Physics 921 South 8th Ave. Stop 8106 Pocatello, ID 83209-8106, USA*

[59]*Imperial College of Science Tech. & Medicine, Blackett Lab.; Prince Consort Road; London SW7 2BZ, UK*

[60]*Indian Institute of Technology Guwahati, Guwahati, 781 039, India*

[61]*Indiana University, Bloomington, IN 47405-7105, USA*

[62]*Iowa State University, Ames, IA 50011, USA*

[63]*Institut de Physique Nucleaire de Lyon (IPNL), Rue E. Fermi 4 69622 Villeurbanne, France*

[64]*University of Jammu, Physics Department, JAMMU-180006, India*

[65]*Kansas State University, Manhattan, KS 66506, USA*

[66]*Kavli IPMU, University of Tokyo, Kashiwa Shi, Chiba 277-8568, Japan*

[67]*KEK, High Energy Accelerator Research Organization 1–1 Oho, Tsukuba-shi; Ibaraki-ken 305-0801, Japan*

[68]*KYIV National University, Department of Nuclear Physics, 64, 01601 Kyiv, Ukraine*

[69]*Koneru Lakshmaiah, India*

[70]*Krakow, Golebia 24, 31-007 Kraków, Poland*

[71]*Lab. d'Annecy-le-Vieux de Phys. des Particules, BP 110; F-74941 Annecy-le-Vieux CEDEX, France*

[72]*Laboratori Nazionali del Gran Sasso, I-67010 Assergi, AQ, Italy*

[73]*Lancaster University, Bailrigg, Lancaster LA1 4YB, UK*

[74]*Lawrence Berkeley National Lab., Berkeley, CA 94720-8153, USA*

[75]*Université de Liège, Bat B5, Sart Tilman B-4000, Belgium*

[76]*University of Liverpool, L69 7ZE, Liverpool, UK*

[77]*University College London, London, WC1E 6BT, UK*

[78]*Los Alamos National Laboratory, Los Alamos, NM 87545, USA*

[79]*Louisiana State University, Baton Rouge, LA 70803-4001, USA*

[80]*University of Lucknow, Department of Physics Lucknow 226007 Uttar Pradesh, India*

[81]*University of Manchester, Oxford Road, Manchester M13 9PL, UK*

[82]*University of Maryland, College Park, MD 20742-4111, USA*

[83]*Massachusetts Institute of Technology, Cambridge, MA 02139-4307, USA*

[84]*Max Planck MPP, Max-Planck-Institut fuer Physik (Werner-Heisenberg-Institut) Foehringer Ring 6 80805 Muenchen, Germany*

[85]*University of Puerto Rico, Box 9016; Mayaguez, PR 00681-9000, USA*

[86]*University of Michigan, 450 Church Street Ann Arbor, Michigan 48109-1040, USA*

[87]*Michigan State University, East Lansing, MI 48824, USA*

[88]*University di Milano, INFN Sezione di Milano, I-20133 Milano, Italy*

[89]*INFN Sezione di Milano Bicocca, Piazza della Scienza 3, 20126 Milano, Italy*

[90]*University of Minnesota (Twin Cities), Minneapolis, MN 55455, USA*

[91]*University of Minnesota (Duluth), Duluth, MN 55812, USA*

[92]*NIKHEF, Science Park, Amsterdam, Netherlands*

[93]*Istituto Nazionale di Fisica Nucleare - Sezione di Napoli, Complesso Universitario di Monte S. Angelo, I-80126 Napoli, Italy*

[94]*National Centre for Nuclear Research, A. Soltana 7, 05 400 Otwock, Poland*

[95]*Jawaharlal Nehru University, New Delhi 110067, INDIA, India*

[96]*University of New Mexico, MSC07 4220; Albuquerque, NM 87131, USA*

[97]*Northwestern University, Evanston, Il 60208, USA*

[98]*University of Notre Dame, Notre Dame, IN 46556-5670, USA*

[99]*Observatorio Nacional, R. Lourival Bpo., 89 - Madruga, Vassouras - RJ, 27700-000, Brazil*

[100]*Ohio State University, Dept. of Physics; 191 W. Woodruff Ave.; Columbus, OH 43210, USA*

[101]*Oregon State University, Dept. of Physics; 301 Weniger Hall; Corvallis, OR 97331–6507, USA*

[102]*University of Oxford, Oxford, OX1 3RH, UK*

[103]*Pacific Northwest National Lab, , USA*

[104]*Pennsylvania State University, PMB 264; University Park, PA 16802-6300, USA*

[105]*PUCP, Av. Universitaria 1801, Lima, Peru*

[106]*University of Padova, Dip. Fisica e Astronomia G. Galilei and INFN Sezione di Padova, I-35131 Padova, Italy*

[107]*Panjab University, Chandigarh, 160014 U.T., India*

[108]*University of Pavia, INFN Sezione di Pavia, I-27100 Pavia, Italy*

[109]*University of Pennsylvania, Philadelphia, PA 19104-6396, USA*

[110]*University di Pisa, Theor. Division; Largo B. Pontecorvo 3, Ed. B-C; I-56127 Pisa, Italy*

[111]*University of Pittsburgh, Pittsburgh, PA 15260, USA*

[112]*Princeton University, Princeton, New Jersey 08544-0708, USA*

[113]*Punjab Agri. University, Centre for High Energy Physics; Lahore - 54590, India*

[114] *University of Rochester, Rochester, NY 14627-0171, USA*

[115] *SLAC National Acceleratory Laboratory, Menlo Park, CA 94025, USA*

[116] *STFC Rutherford Appleton Laboratory, Harwell Oxford, Didcot OX11 0QX, UK*

[117] *CEA/Saclay, IPhT; Inst. de Physique Theorique; Orme des Merisiers, Point Courrier 136; F-91191 Gif-sur-Yvette CEDE, France*

[118] *University of Sheffield, Sheffield, S3 7RH, UK*

[119] *University of Sofia, Atomic Physics Dept., Faculty of Physics; 5 James Bourchier Blvd.; BG-1164 Sofia, Bulgaria*

[120] *University of South Carolina, Columbia, SC 29208, USA*

[121] *University of South Dakota, Vermillion, SD 57069, USA*

[122] *South Dakota School of Mines and Technology, Rapid City, SD 57701, USA*

[123] *South Dakota Science And Technology Authority, Lead, SD 57754, USA*

[124] *South Dakota State University, Brookings, SD 57007, USA*

[125] *Southern Methodist University, Dallas, TX 75275, USA*

[126] *Stanford University, Varian Physics Bldg.; 382 Via Pueblo Mall; Stanford, CA 94305–4060, USA*

[127] *Stony Brook University, Nucleon Decay and Neutrino Physics Group Department of Physics and Astronomy Stony Brook University Stony Brook, NY 11794-3800, USA*

[128] *University of Sussex, Brighton, BN1 9RH, UK*

[129] *Syracuse University, Syracuse, NY 13244-1130, USA*

[130] *University of Tennessee at Knoxville, Knoxville, TN, USA*

[131] *University of Texas (Arlington), Arlington, TX 76019, USA*

[132] *University of Texas (Austin), Austin, TX 78712-0264, USA*

[133] *TUBITAK Space Technologies Research Institute , TR-06800, Ankara, Turkey*

[134] *Tufts University, Medford, MA 02155, USA*

[135] *Variable Energy Cyclotron Centr, 1/AF, Bidhannagar Kolkata - 700 064 West Bengal, India*

[136] *Instituto de Fisica Corpuscular, C/Catedratico Jose Beltran, 2 E-46980 Paterna (Valencia), Spain*

[137] *Virginia Tech., Blacksburg, VA 24061-0435, USA*

[138] *University of Warsaw, Faculty of Physics ul. Pasteura 5 02-093 Warsaw, Poland*

[139] *University of Warwick, Coventry CV4 7AL, UK*

[140] *Wichita State University, Physics Division 1845 Fairmount St. Wichita, KS 67260-0032, USA*

[141] *College of William and Mary, Williamsburg, VA 23187-8795, USA*

[142] *University of Wisconsin, Madison, WI 53706, USA*

[143] *Wroclaw University, Plac Maxa Borna 9, 50-204 Wroclaw, Poland*

[144] *Yale University, New Haven, CT 06520, USA*

[145] *Yerevan Institute for Theoretical Physics and Modeling, Halabian Str. 34; Yerevan 0036, Armenia*

[146] *York, Physics and Astronomy Dept.; 4700 Keele St.; Toronto M3J 1P3, Canada*

[147] *ETH Zurich, HPK F 23 Schafmattstr. 20 8093 Zürich, Switzerland*

2015-11-25



\renewcommand{\familydefault}{\sfdefault}
\renewcommand{\thepage}{\roman{page}}
\setcounter{page}{0}

\pagestyle{plain} 


\setcounter{tocdepth}{2}
\textsf{\tableofcontents}

\textsf{\listoffigures}

\textsf{\listoftables}

\printnomenclature

\iffinal\else
\textsf{\listoftodos}
\clearpage
\fi

\renewcommand{\thepage}{\arabic{page}}
\setcounter{page}{1}

\pagestyle{fancy}

\renewcommand{\chaptermark}[1]{%
\markboth{Chapter \thechapter:\ #1}{}}
\fancyhead{}
\fancyhead[RO,LE]{\textsf{\footnotesize \thechapter--\thepage}}
\fancyhead[LO,RE]{\textsf{\footnotesize \leftmark}}

\fancyfoot{}
\fancyfoot[RO]{\textsf{\footnotesize LBNF/DUNE Conceptual Design Report}}
\fancyfoot[LO]{\textsf{\footnotesize \thedoctitle}}
\fancypagestyle{plain}{}

\renewcommand{\headrule}{\vspace{-4mm}\color[gray]{0.5}{\rule{\headwidth}{0.5pt}}}

%% file: common/acronyms-shared-vol1-2-4.tex
\nomenclature{$\mathcal{O}(n)$}{of order $n$}
\nomenclature{3D}{3 dimensional (also 1D, 2D, etc.)} 
\nomenclature{CDR}{Conceptual Design Report}
\nomenclature{CF}{Conventional Facilities}
\nomenclature{CP}{product of charge and parity transformations}
\nomenclature{CPT}{product of charge, parity and time-reversal transformations}
\nomenclature{CPV}{violation of charge and parity symmetry}
\nomenclature{DAQ}{data acquisition}
\nomenclature{DOE}{U.S. Department of Energy}
\nomenclature{DUNE}{Deep Underground Neutrino Experiment}
\nomenclature{ESH}{Environment, Safety and Health}
\nomenclature{eV}{electron volt, unit of energy (also keV, MeV, GeV, etc.)}
\nomenclature{FD}{far detector}
\nomenclature{FGT}{Fine-Grained Tracker}
\nomenclature{FSCF}{far site conventional facilities}
\nomenclature{NSCF}{near site conventional facilities}
\nomenclature{GUT}{grand unified theory}
\nomenclature{\ktyr}{exposure (without beam), expressed in kilotonnes times years}
\nomenclature{\ktMWyr}{exposure, expressed in kilotonnes $\times$ megawatts $\times$ years, based on 56\% beam uptime and efficiency} 
\nomenclature{L}{level, indicates depth in feet underground at the far site, e.g., 4850L}
\nomenclature{LAr}{liquid argon}
\nomenclature{LArTPC}{liquid argon time-projection chamber}
\nomenclature{LBL}{long-baseline (physics)}
\nomenclature{LBNF}{Long-Baseline Neutrino Facility}
\nomenclature{MH}{mass hierarchy}
\nomenclature{MI}{Main Injector (at Fermilab)}
\nomenclature{ND}{near neutrino detector}
\nomenclature{NDS}{Near Detector Systems; refers to the collection of detector systems at the near site }
\nomenclature{near detector}{except in Volume 4 Chapter 7, \textit{near detector} refers to the \textit{neutrino} detector system in the NDS}
\nomenclature{POT}{protons on target}
\nomenclature{QA}{quality assurance}
\nomenclature{SM}{Standard Model of particle physics}
\nomenclature{t}{metric ton, written \textit{tonne} (also kt)}
\nomenclature{tonne}{metric ton}
\nomenclature{TPC}{time-projection chamber (not used as `total project cost' in the CDR)}

%% file: common/acronyms-shared-vol1-4.tex

\nomenclature{APA}{anode plane assembly} 
\nomenclature{BLM}{(in Volume 4) beamline measurement (system); (in Volume 3) beam loss monitor}
\nomenclature{CPA}{cathode plane assembly}
\nomenclature{ECAL}{electromagnetic calorimeter}
\nomenclature{GAr}{gaseous argon}
\nomenclature{HV}{high voltage}

%% file: common/acronyms-shared-vol1-2.tex
\nomenclature{CKM}{(CKM matrix) Cabibbo-Kobayashi-Maskawa matrix, also known as
quark mixing matrix} 
\nomenclature{C.L.}{confidence level}
\nomenclature{octant}{any of the eight parts into which 4$\pi$ is divided by three mutually perpendicular axes; the range of the PMNS angles is $0$ to $\pi/2$, which spans only two of the eight octants}
\nomenclature{PMNS}{(PMNS matrix) Pontecorvo-Maki-Nakagawa-Sakata matrix, also known as
the lepton or neutrino mixing matrix} 

%% file: common/acronyms-vol1.tex
\nomenclature{L1, L2, ...}{WBS level within the LBNF and DUNE Projects, where the overall Project is L1}
\nomenclature{MOU}{memorandum of understanding}
\nomenclature{PIP-II(III)}{Proton Improvement Plan (II or III)}
\nomenclature{SDSTA}{South Dakota Science and Technology Authority}
\nomenclature{WBS}{Work Breakdown Structure}

%% file: volume-project/chapter-overview.tex
\chapter{Introduction to LBNF and DUNE}
\label{ch:project-overview}

%
%
\section{International Convergence}

During the last decade, several independent worldwide efforts have attempted to develop paths towards a next-generation long-baseline neutrino experiment, including in the U.S. with LBNE, in Europe with LBNO and in Japan with Hyper-Kamiokande. The community has 
generally recognized that putting in place 
the conditions necessary to 
execute this challenging science program in a comprehensive way requires previously independent 
efforts to converge. 

In this context, the Deep Underground Neutrino Experiment (DUNE) represents the convergence of a substantial fraction of the worldwide neutrino-physics community around the 
opportunity provided by the 
large investment planned by the U.S. Department of Energy (DOE) to support 
a significant expansion of the underground infrastructure at the Sanford Underground Research 
Facility (SURF) in South Dakota, \SI{1300}{\km} from Fermilab, and to create a megawatt neutrino-beam facility at Fermilab by 2026.  The PIP-II accelerator upgrade~\cite{pip2-2013} at 
Fermilab will drive the new neutrino beamline at Fermilab with a beam power\footnote{assuming a \SI{120}{\GeV} primary proton beam. For a \SI{80}{\GeV} primary proton beam, the corresponding beam power is \SI{1.07}{\MW}.} of up to \SI{1.2}{\MW}, with a planned upgrade 
of the accelerator complex to enable it to provide up to \SI{2.4}{\MW} of beam power by 2030.  

This document presents 
the Conceptual Design Report (CDR) put forward by an international neutrino community to pursue 
the Deep Underground Neutrino Experiment at the Long-Baseline Neutrino Facility (LBNF/DUNE),
a groundbreaking science experiment for long-baseline neutrino oscillation studies and for neutrino astrophysics and nucleon decay searches. The DUNE far detector will be a very large modular liquid argon time-projection chamber (LArTPC) located deep underground, coupled to the LBNF multi-megawatt  
wide-band neutrino beam.   DUNE will also have a high-resolution and high-precision near detector.

The physics case for the LBNF neutrino facility was highlighted as a strategic priority in the 2014 P5 report~\cite{p5report2014}.
P5 identified the following minimum requirements for LBNF to proceed: 
the identified capability to reach an exposure of at least 120~\ktMWyr{}~\footnote{An exposure
of 1 MW.year corresponds to $1\times 10^{21}$ protons-on-target per year at 120 GeV. This includes the LBNF beamline efficiency which is estimated to be 56\%.}  by the 2035 timeframe;
the far detector situated underground with cavern space for expansion to at least 40-kt LAr fiducial;
1.2-MW beam power upgradable to multi-megawatt power;
demonstrated capability to search for supernova bursts; and
a demonstrated capability to search for proton decay, 
providing a significant improvement in discovery sensitivity over current searches for the proton lifetime.
Furthermore, P5 identified  the \textit{goal} of a sensitivity to CP violation of better than 3$\sigma$ over more than 75\% 
of the range of possible values of the unknown CP-violating phase \deltacp.
The strategy presented in this CDR meets all of these requirements.

\section{The LBNF/DUNE Conceptual Design Report Volumes}

\subsection{A Roadmap of the CDR}

The LBNF/DUNE CDR describes the proposed physics program and 
conceptual technical designs of the facility and detectors.  At this stage, the design is
still undergoing development and the CDR therefore presents a \textit{reference design} for each element as well as any 
\textit{alternative designs} that are under consideration.

The CDR is composed of four volumes and is supplemented
by several annexes that provide details of the physics program and technical designs. The volumes are as follows:

\begin{itemize}
\item \volintro{} --- provides an executive summary of and strategy for the experimental 
program and of the CDR as a whole.
\item \href{http://arxiv.org/abs/1512.06148}{\volphys{}} --- outlines the scientific objectives and describes the physics studies that 
the DUNE collaboration will undertake to address them.
\item \vollbnf{} --- describes the LBNF project, which includes design and construction of the 
beamline at Fermilab, the conventional facilities at both Fermilab and SURF, and the cryostat
 and cryogenics infrastructure required for the DUNE far detector.
\item \href{http://arxiv.org/abs/1601.02984}{\voldune{}} --- describes the DUNE project, which includes the design, construction and 
commissioning of the near and far detectors. 
\end{itemize}

More detailed information for each of these volumes is provided in a set of annexes listed on the \href{https://web.fnal.gov/project/LBNF/SitePages/Proposals%20and%20Design%20Reports.aspx}{Proposals and Design Reports} 
page.

\subsection{About this Volume}

This introductory volume of the LBNF/DUNE Conceptual Design Report provides an overview of LBNF and
DUNE (Chapter~\ref{ch:project-overview}), including the strategy that is being developed to construct, install and commission the technical and conventional facilities in accordance with the requirements set out by the P5 report of 2014~\cite{p5report2014}, which, in turn, is in line with the CERN
European Strategy for Particle Physics (ESPP) of 2013~\cite{ESPP-2012}. This volume also introduces the DUNE science program (Chapter~\ref{v1ch:science}) and the technical designs of the facilities and the detectors 
(Chapter~\ref{v1ch:tech-designs}). It concludes with a description of the LBNF and DUNE organization and management structures (Chapter~\ref{v1ch:org-mgmt}).

\section{A Compelling Scientific Program}

The study of the properties of neutrinos has produced 
many surprises, including the evidence for physics beyond the Standard Model of elementary particles and interactions.   The phenomenon of neutrino flavor oscillations, whereby 
neutrinos can transform into a different flavor after traveling a distance, 
is now well established. Important conclusions that follow from these discoveries include that neutrinos have mass and that their 
mass eigenstates are mixtures of their 
flavor eigenstates.

Speculations on the origin of neutrino masses and mixings are wide-ranging. 
Solving the puzzle will require more precise and detailed experimental information with neutrinos and antineutrinos and with sensitivity to matter effects. With the exception of a few anomalous results, the current data can be described in terms of the three-neutrino paradigm, in which the 
quantum-mechanical mixing of the three mass eigenstates produces the three known neutrino-flavor states.  The mixings are described by the Pontecorvo-Maki-Nakagawa-Sakata (PMNS) matrix, a parameterization that includes a CP-violating phase. 

The primary science objectives 
of DUNE are to carry out a comprehensive investigation of neutrino oscillations to test CP violation in the lepton sector, determine the ordering of the neutrino masses, and to test the three-neutrino paradigm.
By measuring \textit{independently} the  propagation of neutrinos and antineutrinos through matter, DUNE will be able to observe 
neutrino transitions with the precision required to determine the 
CP-violating phase and 
the neutrino mass hierarchy.

The construction of LBNF and DUNE will also enable a high-priority ancillary science program, such as 
very precise measurements of neutrino interactions and cross-sections, studies of nuclear effects in such interactions, measurements of the structure of nucleons, as well as precise tests of the electroweak theory. 
These measurements of the properties of neutrino interactions are also necessary 
to achieve the best sensitivities in the long-baseline neutrino oscillation program. 

The DUNE far detector, consisting of four LArTPC modules located deep underground, each with a mass forty times 
larger than ever before built,  
will offer unique capabilities for addressing 
non-accelerator physics topics. These include measuring atmospheric neutrinos, searching for nucleon decay, and measuring astrophysical neutrinos --- possibly even 
the neutrino burst 
from a core-collapse supernova. 
Observations of these kinds will bring new insight into these fascinating natural phenomena. 


An intriguing 
conjecture is that of neutrino masses being related to an 
ultra-high-energy scale that may be associated with the unification of matter and forces. Such theories are able to describe the absence of antimatter in the universe in terms of the properties of ultra-heavy particles; they also 
offer an explanation 
of cosmological inflation in terms of the phase transitions associated with the breaking of symmetries at this ultra-high-energy scale. DUNE's capability to detect and study rare events such as nucleon decays in an unbiased and unprecedented way will allow it to probe these very high-energy scales. 



Finally, further developments of LArTPC 
technology during the course of the DUNE far detector construction may open up the opportunity
to observe very low-energy phenomena such as solar neutrinos or even the diffuse supernova neutrino flux.


\section{Overall LBNF/DUNE Project Strategy} 

The LBNF/DUNE project (the ``project'') strategy presented in this CDR has been developed to meet the requirements 
set out in the P5 report and 
takes into account the recommendations of the CERN European Strategy for Particle 
Physics (ESPP) of 2013, which classified the long-baseline neutrino program as 
one of the four scientific objectives with required international infrastructure.

The Report of the Particle Physics Project Prioritization Panel (P5) 
states that for a long-baseline neutrino oscillation experiment, ``The 
minimum requirements to proceed are the identified capability to reach an exposure 
of \num{120}~\ktMWyr{} by the 2035 timeframe, the far detector situated underground 
with cavern space for expansion to at least 40~kt LAr fiducial volume, and 1.2~MW 
beam power upgradable to multi-megawatt power. The experiment should have the demonstrated 
capability to search for supernova bursts and for proton decay, providing a significant 
improvement in discovery sensitivity over current searches for the proton lifetime.'' 
Based on the resource-loaded schedules for the reference designs of the facility (\vollbnf)
and the detectors (\voldune), the strategy presented here meets these criteria. 

With the availability of space for expansion and improved access at SURF, 
the international DUNE collaboration proposes to construct a deep-underground neutrino observatory based on four independent \ktadj{10} LArTPCs at this site. 
The goal is the deployment of two \ktadj{10} fiducial mass detectors in a relatively short timeframe, followed by future expansion to the full detector size as soon thereafter as possible. 

Several LArTPC designs are under development by different groups worldwide, involving both single- and dual-phase readout technology.
The DUNE 
collaboration has the necessary scientific and technical expertise, 
and international participation  to design and implement this exciting discovery experiment. 

The Long-Baseline Neutrino Facility (LBNF) provides

\begin{itemize}

\item  the  technical and conventional facilities for a powerful \MWadj{1.2} neutrino beam utilizing the PIP-II upgrade of the Fermilab accelerator 
complex, to become operational by 2025 
at the latest, and to be upgradable to \SI{2.4}{\MW} with the proposed 
PIP-III upgrade;

\item  the civil construction (conventional facilities or CF) for the near detector systems at Fermilab; 

\item the excavation of four underground caverns at SURF, planned to be completed 
by 2021 
under a single contract, with each cavern to be capable of housing a cryostat for
a minimum \ktadj{10} fiducial mass LArTPC; and

\item surface, shaft, and underground infrastructure to support 
the outfitting of the caverns with four free-standing, steel-supported cryostats 
and the required cryogenics systems. The first cryostat will be available for filling, after installation of the detector components, by
2023, enabling a rapid deployment of the first two \ktadj{10} far detector modules. 
The intention is to install the third and fourth cryostats as rapidly as funding will 
allow.

\end{itemize}

The Deep Underground Neutrino Experiment (DUNE) provides
\begin{itemize}

\item four massive LArTPCs, each with a fiducial mass of at least \SI{10}{\kt}. The division of 
the far detector into four equal-mass detectors provides the project flexibility 
in the installation and funding (DOE vs. non-DOE); this division also mitigates risks and allows for an early and graded science return.

\item the near detector systems, consisting of a high-resolution neutrino detector 
and the muon monitoring system that will enable 
the precision 
needed to fully 
exploit the statistical power of the 
far detector coupled to the 
MW-class 
neutrino beam.
\end{itemize}

Based on the reference design described below and in Volumes 2, 3 and 4 of this 
CDR, the resource-loaded schedule 
plans for the first two \ktadj{10} far detector modules to be
operational by 2025,
with first beam shortly afterward. 
At that time, the cavern 
space for all four \ktadj{10} far detector modules will be available, allowing for 
an accelerated installation schedule if sufficient funding sources for
the experiment can be established on an accelerated timescale.  

\vspace{6pt}
The project strategy described above meets the experiment's scientific objectives,
 reaching an exposure of 
\num{120}~\ktMWyr{} by 2032, and potentially earlier if additional resources are identified. 
The P5 recommendation of sensitivity to CP violation of 3$\sigma$ for 75\% of $\delta_\text{CP}$
values can be reached with an exposure of \num{850}~\ktMWyr{} with an optimized beam.

\section{The International Organization and Responsibilities}

The 
model used by CERN for managing the construction and exploitation of the LHC and its experiments was used as a starting point for the joint management of LBNF and the experimental program.  Fermilab, as the host laboratory, has the responsibility for the facilities and their operations 
and oversight of the experiment and its operations.  Mechanisms to ensure input from and coordination among all of the funding agencies supporting the collaboration, modelled on the CERN Resource Review Board, have been adopted. 
A similar structure is employed to coordinate among funding agencies supporting the LBNF construction and operation.  

The LBNF/DUNE project will be organized as two distinct entities. The LBNF portion is funded primarily
by the U.S. DOE acting on behalf of the hosting country.  CERN provides in-kind contributions to the LBNF infrastructure needed for the DUNE experiment. The DUNE portion is organized
as an international collaboration; it is adopting a model in which the DOE and international funding agencies share costs 
for the DUNE detectors.

The DUNE collaboration is responsible for
\begin{itemize}
\item the definition of the scientific goals and corresponding scientific and technical requirements on the detector systems and neutrino beamline;
\item the design, construction, commissioning and operation of the detectors; and
\item the scientific research program conducted with the DUNE detectors. 
\end{itemize}

The high-intensity proton source at Fermilab that will drive the long-baseline neutrino beam utilizes the existing 
Main Injector with upgraded injectors (PIP-II).  PIP-II is also being planned with significant international collaboration.  Fermilab, working 
with the participation and support of international partners, is responsible for 
LBNF, including
\begin{itemize}
\item design, construction and operation of the LBNF beamline, including the primary proton beamline and the neutrino beamline including target, focusing structure (horns), decay pipe, absorber, and corresponding beam instrumentation;
\item design, construction and operation of the CF and 
experiment infrastructure on the Fermilab site required for the near detector system; and
\item design, construction and operation of the CF and 
experiment infrastructure 
at SURF, including the cryostats and cryogenics systems, required for the far detector.
\end{itemize}

\section{A Two-Pronged Schedule} 

The schedule for the design and construction work for LBNF and DUNE has two critical parallel paths: one for the 
far site (SURF) and 
another for the 
near site (Fermilab). The schedule for the initial work is driven by the CF design and construction at each site. A summary of the schedule is shown in Figure~\ref{fig:summary-sched}.

\begin{cdrfigure}[High-level summary of LBNF/DUNE schedule]{summary-sched}{High-level summary of LBNF/DUNE schedule}
\includegraphics[width=1.2\textwidth, angle=90]{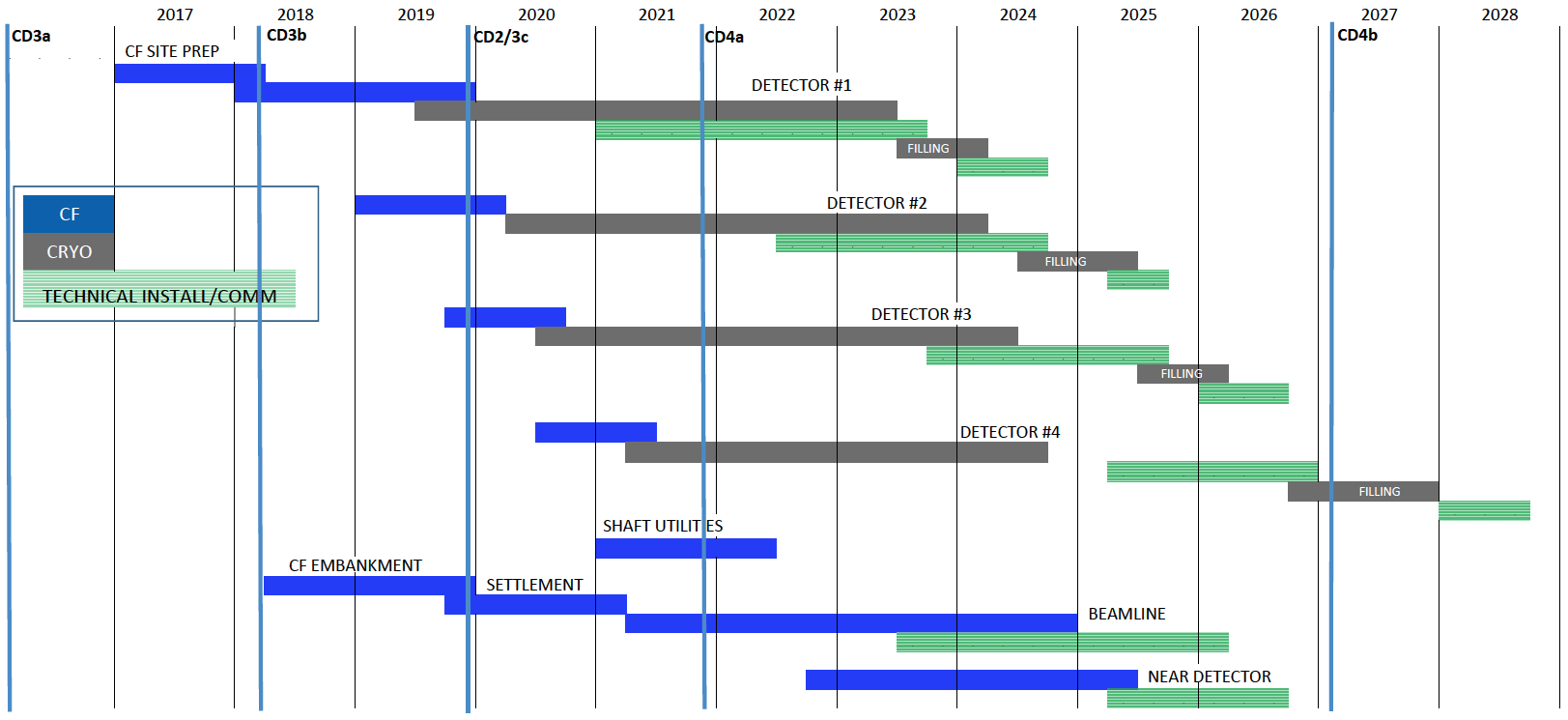}
\end{cdrfigure}

Within the anticipated DOE funding profile, in particular during the initial phase of the project, the far site conventional
facilities are advanced first; their final design starts in fall 2015. Early site preparation is timed to be completed 
in time to start excavation when the Ross Shaft rehabilitation work finishes
 in late 2017. As each detector 
 cavern is excavated and sufficient utilities are installed, the cryostat and cryogenics system work proceeds, followed by detector installation, filling and commissioning. 
 The far detector module \#1 is to be operational by 2024 with modules \#2 and \#3 completed
 one and two years later, respectively, and module \#4 completed by early 2027.

The near site work is delayed with respect to the far site due to the anticipated funding profile. The near site CF and beamline work essentially slows to nearly a stop 
until design restarts in late 2017. Optimization decisions about the beamline that affect the CF design will need to be made by late 2018 in order to be ready for the CF design process. The embankment is constructed and then allowed to settle for at least twelve months before the majority of the beamline CF work proceeds. The beneficial occupancies of the various beamline facilities 
are staggered to allow beamline installation to begin as soon as possible. With this timescale, the far detector science program 
starts with the first module installed and no beam, focusing on non-accelerator-based science 
for slightly more than one year until 
the beamline installation is completed.

The near detector CF construction overlaps that for the beamline, but lags due to available funding. The near detector assembly begins on the surface before beneficial occupancy, after which the detector is installed, complete at about the same time as far detector module \#4. 

The DOE project management process requires approvals at Critical Decision milestones, which allow the LBNF/DUNE project to move to the next step. In fall 2015 the far site CF will seek CD-3a approval for construction of some of the CF and cryogenics systems at SURF. In spring 2018 LBNF near site CF will seek CD-3b construction approval for Advanced Site Preparation to build the embankment. In 2020 LBNF and DUNE will seek to baseline the LBNF/DUNE scope of work, cost and schedule, as well as construction approval for the balance of the project scope of work. 
The project concludes with CD-4 approval to start operations.

%% file: volume-project/chapter-science.tex
\chapter{DUNE Science}
\label{v1ch:science}

DUNE will address fundamental questions key to our understanding of the universe. These include:
\begin{itemize}
   \item {\bf What is the origin of the matter-antimatter asymmetry in the universe?} Immediately after
                    the Big Bang, matter and antimatter were created equally, but now matter dominates.
                    By studying the properties of neutrino and antineutrino oscillations, LBNF/DUNE 
                    will pursue the current most promising avenue for understanding this asymmetry.
   \item {\bf What are the fundamental underlying symmetries of the universe?} The patterns of mixings and masses between the particles of the Standard Model is not understood. By making precise measurements of the mixing between the neutrinos and the ordering of neutrino masses and comparing these with the quark sector, LNBF/DUNE could reveal new underlying symmetries of the universe.
  \item{\bf  Is there a Grand Unified Theory of the Universe?} Results from a range of experiments suggest that the
                 physical forces observed today were unified into one force at the birth of the universe.
                Grand Unified Theories (GUTs), which attempt to describe the unification of forces,
                predict that protons should decay, a process that has never been observed. DUNE will 
                search for proton decay in the range of proton lifetimes predicted by a wide range of GUT models.
   \item{\bf How do supernovae explode and what new physics will we learn from a neutrino burst?}
   Many of the heavy elements that are the key components of life were created in the super-hot cores of collapsing stars. DUNE would be able to detect the neutrino bursts from core-collapse supernovae within our galaxy (should any occur). Measurements of the time, flavor and energy structure of the neutrino burst will be critical for understanding the dynamics of this important astrophysical phenomenon, as well as bringing information on neutrino properties and other particle physics.
\end{itemize}

\section{DUNE Scientific Objectives}

The DUNE scientific objectives are categorized into: the \textit{primary science program}, addressing the key science questions 
highlighted by the particle physics project prioritization panel (P5); 
a high-priority \textit{ancillary science program} that is 
enabled by the construction of LBNF and DUNE; and \textit{additional scientific objectives}, that may require further developments 
of the LArTPC technology. A detailed description of the physics objectives of DUNE is provided in \href{http://arxiv.org/abs/1512.06148}{Volume 2 of the CDR}.

\subsection{The Primary Science Program}

The primary science program of LBNF/DUNE  focuses on fundamental open questions in neutrino and astroparticle physics: 
\begin{itemize}
  \item precision measurements of the parameters that govern $\nu_{\mu} \rightarrow \nu_\text{e}$ and
           $\overline{\nu}_{\mu} \rightarrow \overline{\nu}_\text{e}$ oscillations with the goal of
  \subitem -- measuring the charge-parity (CP) violating phase $\delta_\text{CP}$, where a value differing from zero or $\pi$ would represent the discovery of CP violation in the leptonic sector, providing a possible explanation for the matter-antimatter asymmetry in the universe;
  \subitem -- determining the neutrino mass ordering (the sign of $\Delta m^2_{31} \equiv m_3^2-m_1^2$), often referred to as the neutrino \textit{mass hierarchy}; and
  \subitem -- precision tests of the three-flavor neutrino oscillation paradigm through studies of muon neutrino disappearance 
    and electron neutrino appearance in both $\nu_\mu$ and $\overline{\nu}_{\mu}$ beams, including the 
    measurement of the mixing angle $\theta_{23}$ and the determination of the octant in which this angle lies.
    \item search for proton decay in several important decay modes, for example $\text{p}\rightarrow\text{K}^+\overline{\nu}$, where the observation of proton decay would represent a ground-breaking discovery in physics, providing a portal to Grand Unification of the forces; and
    \item detection and measurement of the $\nu_\text{e}$ flux from a core-collapse supernova within our galaxy, should one occur during the lifetime of the DUNE experiment.
\end{itemize}

\subsection{The Ancillary Science Program}

The intense neutrino beam from LBNF, the massive DUNE LArTPC far detector and the high-resolution
  DUNE near detector provide a rich ancillary science program, beyond the primary mission of the experiment. The ancillary science program includes
\begin{itemize}
     \item other accelerator-based neutrino flavor transition measurements with sensitivity to Beyond Standard Model (BSM) physics, such as: non-standard interactions (NSIs); the search for sterile neutrinos at both the near and far sites;
 and measurements of tau neutrino appearance;
     \item measurements of neutrino oscillation phenomena using atmospheric neutrinos;
     \item a rich neutrino interaction physics program utilizing the DUNE near detector, including: a wide-range of measurements of neutrino cross sections; studies of nuclear effects, including neutrino final-state interactions; measurements of the structure of nucleons; and  measurement of $\sin^2\theta_\text{W}$; and
     \item  the search for signatures of dark matter.
\end{itemize} 
Furthermore, a number of previous breakthroughs in particle physics have been serendipitous, in the sense that they were beyond the
original scientific objectives of an experiment. The intense LBNF neutrino beam and novel capabilities for both 
the DUNE near and far detectors will probe new regions of parameter space for both the accelerator-based and astrophysical frontiers, 
providing the opportunity for discoveries that are not currently anticipated.

\section{Long-Baseline Neutrino Oscillation Physics}

Precision neutrino oscillation measurements lie at the heart of the DUNE scientific program.
The \SIadj{1300}{\km} baseline, coupled with the wide-band
high-intensity neutrino beam from LBNF, establishes one of DUNE's key
strengths, namely sensitivity to the matter effect. This effect leads to a
discrete asymmetry in the \numu $\to$ \nue versus \anumu $\to$ \anue
oscillation probabilities, the sign of which depends on the presently
unknown mass hierarchy (MH).  At \SI{1300}{\km} the asymmetry,
\begin{equation}
\mathcal{A} = \frac{ P(\nu_\mu \rightarrow \nu_e)-P(\bar{\nu}_\mu \rightarrow \bar{\nu}_e)}{P(\nu_\mu \rightarrow \nu_e)+P(\bar{\nu}_\mu \rightarrow \bar{\nu}_e)}
\end{equation}
is approximately $\pm 40\%$ in the region of the peak flux in the
absence of CP-violating effects. This is larger than the maximal
possible CP-violating asymmetry associated with the CP-violating
phase, \deltacp, of the three-flavor PMNS mixing matrix in the region of
the peak flux. The CP asymmetry is larger in the energy regions below the peak
flux while the matter asymmetry is smaller. As a result, the LBNF
wide-band beam will allow unambiguous determination of both the MH and
\deltacp with high confidence \textit{within the same experiment}, i.e., DUNE. 
The DUNE science reach is described in detail in \volphys, where it is presented 
in terms of  exposure expressed in units of \ktMWyr{}.
For instance, seven years of data
(\num{3.5} years in neutrino mode plus \num{3.5} years in antineutrino
mode\footnote{unless otherwise stated, the results presented in the CDR assume equal running in neutrino and antineutrino mode.}) with a \ktadj{40} detector and a \MWadj{1.07} beam (based on a \GeVadj{80} primary proton beam) correspond to an
exposure of \SI{300}~\ktMWyr. 

The DUNE far detector will be built as four \ktadj{10} modules, which will
come online sequentially over the course of several years, as described in Chapter~\ref{ch:project-overview}. 
This staged program enables an early scientific output from DUNE, 
initially focused on the observation of natural
sources of neutrinos, searches for nucleon decays and 
measurements of backgrounds. 
About a year after commissioning the first detector module, 
the LBNF neutrino
beam at Fermilab will 
begin sending neutrinos over the \kmadj{1300}
baseline, commencing the LBL oscillation physics program with a beam power of up to \SI{1.2}\MW{}. 
Prior to the operation of the near detector (ND), which
is likely to start after the initial beam running, the early physics program
will be statistically limited. However, the constraints from comparison of the $\nu_\mu$
disappearance spectrum with that from $\nu_e$ appearance mitigate, in part,
the absence of a direct flux measurement from the ND. Subsequently, the ND
measurements will provide powerful constraints on the beam flux, providing the
necessary control of systematic uncertainties for the full exploitation of LBNF/DUNE. 


The evolution of the projected DUNE sensitivities as a function of real time
(for the first \num{15} years of operation) was estimated based on an assumed deployment plan
with the following assumptions:
\begin{itemize}
\item Year 1: \SI{10}\kt{} far detector mass, \MWadj{1.07} \GeVadj{80}
  proton beam with $1.47 \times 10^{21}$ protons-on-target per year
  and no ND
\item Year 2: Addition of the second \ktadj{10} far detector module, for a total far detector mass of
  \SI{20}\kt
\item Year 3: Addition of the third \ktadj{10} far detector module, for a total far detector mass of
  \SI{30}\kt, and first constraints from the preliminary ND data analysis
\item Year 4: Addition of the fourth \ktadj{10} far detector module, for a total far detector mass of
  \SI{40}\kt
\item Year 5: Inclusion of constraints from a full ND data analysis
 \item Year 7: Upgrade of beam power to \SI{2.14}{\MW} for a \SIadj{80}{\GeV}
  proton beam
\end{itemize}
The staging of the detectors and facility in the resource-loaded schedule leads to a similar
evolution of physics sensitivity as a function of time.
In addition, it was assumed that the knowledge from the near detector can be
retroactively applied to previous data sets, such that each
improvement in the knowledge of systematic uncertainties~\footnote{A
  detailed discussion of the systematic uncertainties assumed, given a
  near detector, is presented in \volphys. For studies without a near
  detector an uncertainty of 10\% is assumed on the unoscillated flux
  at the far detector based on the current performance of the NuMI
  beam simulation, with uncertainties on physics backgrounds $\geq
  10\%$ depending on the background.} is applied to the full exposure
up to that point.


%

The discriminating power between the two MH hypotheses is quantified
by the difference, denoted $\Delta \chi^2$, between the
$-2\log{\cal L}$ values calculated for the normal and inverted
hierarchies. As the sensitivity depends on the true value of the unknown
CP-violating phase, \deltacp, all possible values of \deltacp are
considered\footnote{For the case of the MH determination, the usual
  association of this test statistic with a $\chi^2$ distribution for
  one degree of freedom is incorrect; additionally the assumption of a
  Gaussian probability density 
  implicit in this notation is not exact.  The discussion in Chapter~3
  of \volphys{} provides a brief description of the statistical
  considerations.}.  In terms of this test statistic, the MH
sensitivity of DUNE with an exposure of \SI{300}~\ktMWyr{} is
illustrated in Figure~\ref{fig:mhexec} for the case of normal
hierarchy and the current best-fit value of \sinst{23} = 0.45. 
For this exposure, the DUNE determination of the MH will be definitive for
the overwhelming majority of the  \deltacp and \sinst{23} parameter space.
Even for unfavorable combinations of the parameters, a statistically
ambiguous outcome is highly unlikely.  
\begin{cdrfigure}[Summary of mass hierarchy sensitivities]{mhexec}{The
    square root of the mass hierarchy discrimination metric $\Delta
    \chi^2$ is plotted as a function of the unknown value of \deltacp
    for an exposure of \SI{300}~\ktMWyr{} 
    (left).  The minimum significance
    --- the lowest point on the curve on the left --- with which the mass
    hierarchy can be determined for all values of \deltacp as a
    function of years of running under the staging plan described in the text (right).
    The shaded regions represent the range in sensitivity corresponding to
    the different beam design parameters.}
\includegraphics[width=0.49\textwidth]{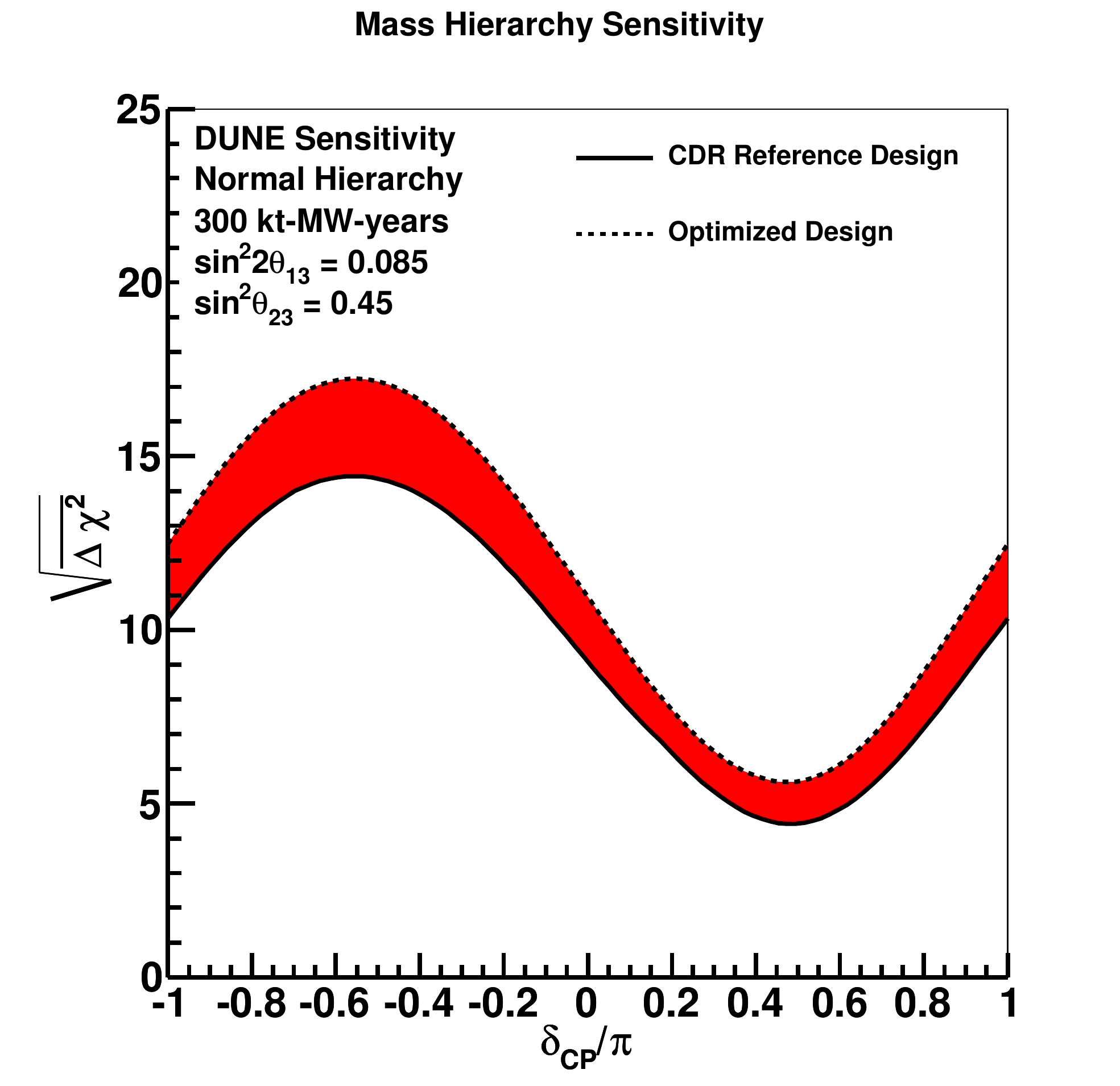}
\includegraphics[width=0.49\textwidth]{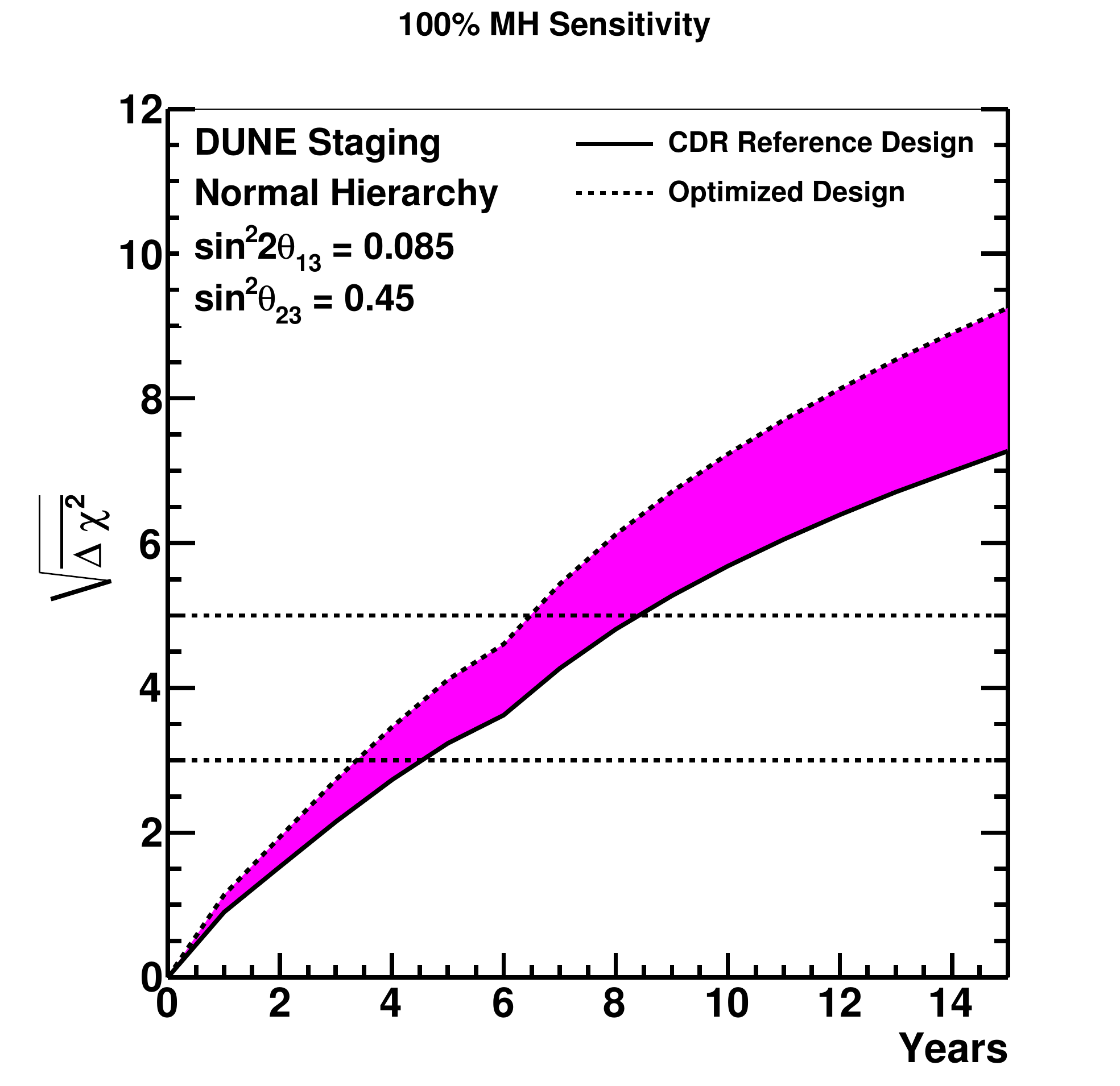}
\label{fig:mhexec}
\end{cdrfigure}

Figure~\ref{fig:mhexec} shows the evolution of the sensitivity to the MH determination as a function
of years of operation, for the least favorable scenario, corresponding to the case in which the MH asymmetry is
maximally offset by the leptonic CP asymmetry. For the reference design beam an exposure of \SI{400}~\ktMWyr{}  
(which corresponds to \num{8.5} years of operation) 
is required to distinguish
between normal and inverted hierarchy with $|\Delta \chi^2| =
\overline{|\Delta \chi^2|} = 25$.  This corresponds to a $\geq
99.9996\%$ probability of determining the correct hierarchy. 
Investments in a more capable target and horn focusing system can
lower the exposure needed to reach this level of sensitivity from
\SI{400}~\ktMWyr{} to around \SI{230}~\ktMWyr{} (\num{6.5} years of
running in the example staging plan). The dependence of the mass
hierarchy sensitivity on systematics is still under evaluation, but
current studies indicate a only weak dependence on the assumptions for 
the achievable systematic uncertainties. This indicates that a measurement of the unknown
neutrino mass hierarchy with very high precision can be carried out
during the first few years of operation with an optimized beamline
design, discussed in Section~\ref{v1ch:tech-designs:beam-strategy} and \vollbnf.
Concurrent analysis of the corresponding atmospheric-neutrino
samples in an underground detector will improve the precision and
speed with which the MH is resolved.

DUNE will search for CP violation using the \numu to \nue and \anumu
to \anue oscillation channels, with two objectives.  First, DUNE aims
to observe a signal for leptonic CP violation independent of the
underlying nature of neutrino oscillation phenomenology. Such a signal
will be observable in comparisons of $\nu_\mu \rightarrow \nu_e$ and
$\bar{\nu}_{\mu} \rightarrow \bar{\nu}_e$ oscillations of the LBNF
beam neutrinos in a wide range of neutrino energies over the
\SIadj{1300}{\km} baseline.
Second,
DUNE aims to make a precise determination of the value of \deltacp
within the context of the standard three-flavor mixing scenario
described by the PMNS neutrino mixing matrix. Together, the pursuit of
these two goals provides a thorough test of the standard three-flavor
scenario.
%
\begin{cdrfigure}[CP-violation sensitivity and $\delta_{\rm CP}$
  resolution as a function of exposure]{execsummaryCP}{The
    significance with which CP violation can be determined for 75\% of
    \deltacp values (left) and the expected 1$\sigma$ resolution
    (right) as a function of exposure in years using the proposed
    staging plan outlined in this chapter. The shaded regions
    represent the range in sensitivity due to potential variations in
    the beam design. The plots assume normal mass hierarchy.}
\includegraphics[width=0.49\textwidth]{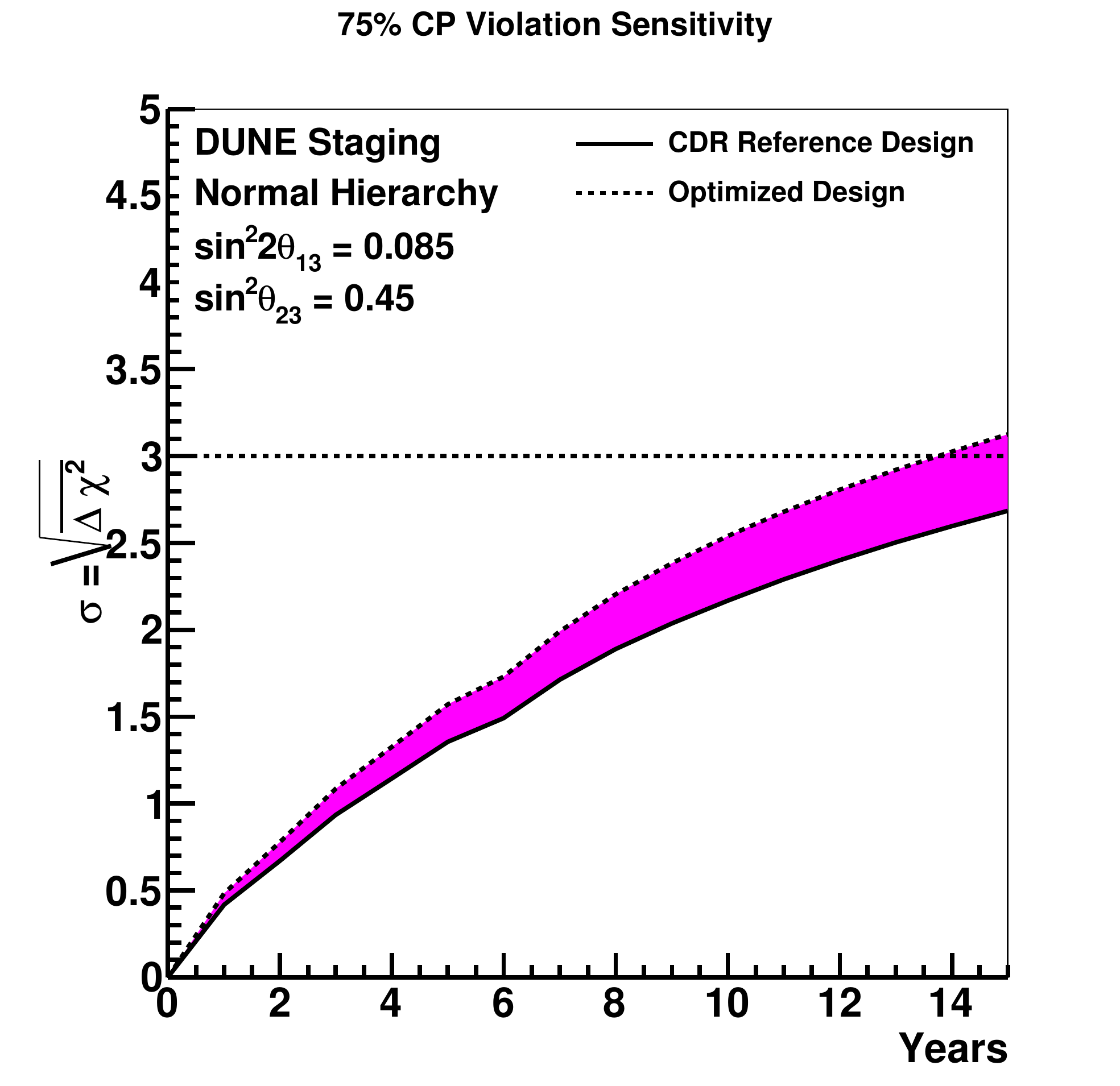}
 \includegraphics[width=0.49\textwidth]{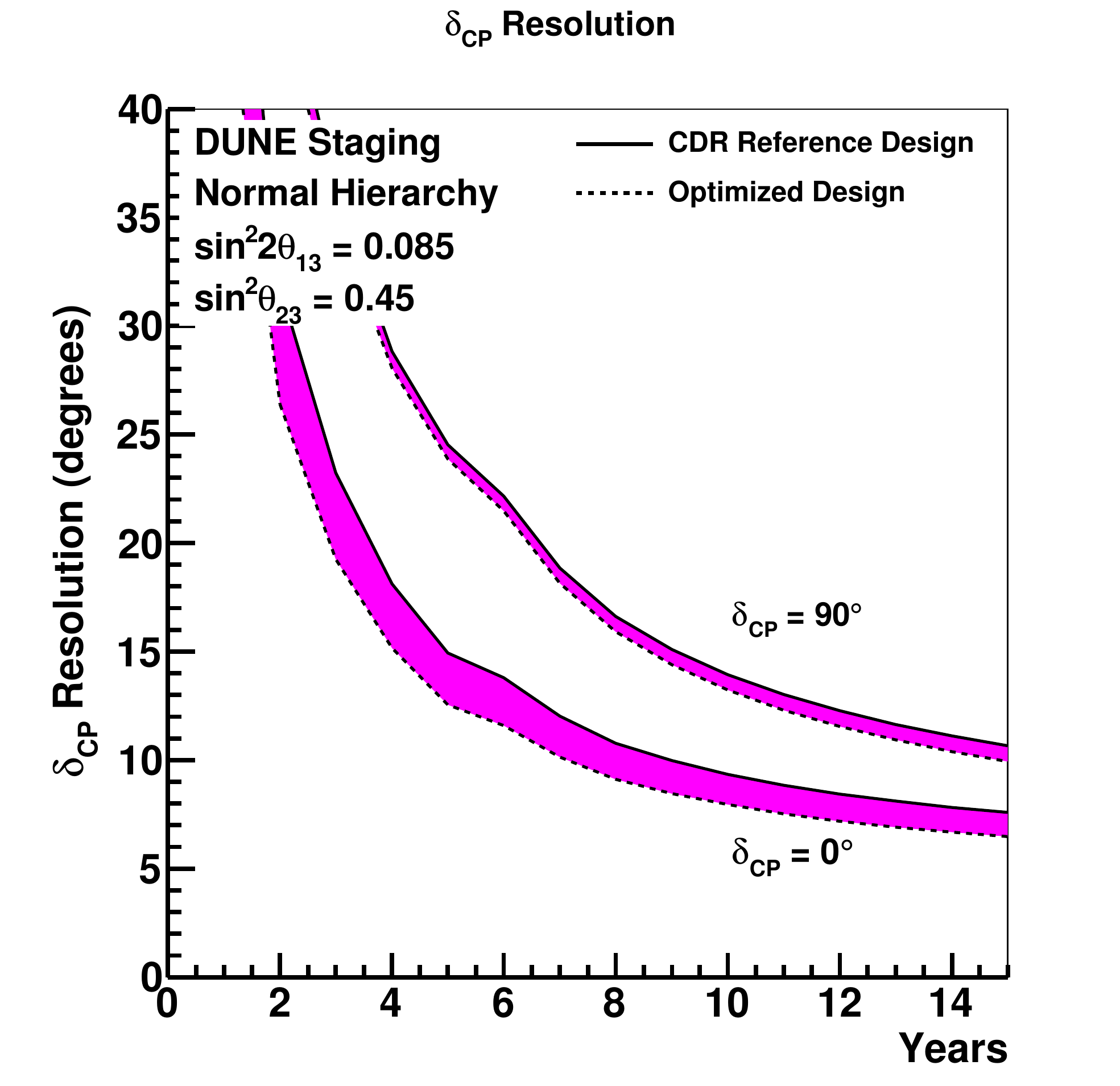}
\end{cdrfigure}
Figure~\ref{fig:execsummaryCP} shows, as a function of time, the
expected sensitivity to CP violation expressed as the minimum significance
with which CP violation can be determined for 75\% of
\deltacp values.
Also shown is the 1$\sigma$ resolution for \deltacp as a
function of time for $\delta_{\rm CP}=0$ (no CP violation) and
$\delta_{\rm CP}=90^\circ$ (maximal CP violation). In both figures the staging scenario
described above was assumed.  The exposure required to measure
$\delta_{\rm CP} = 0 $ with a precision better than $10^\circ$ ranges
from 290 to \SI{450}~\ktMWyr{} depending on the beam design.
A full-scope LBNF/DUNE operating with multi-megawatt 
beam power can eventually achieve a precision 
comparable to the current precision on the CP phase in the
CKM matrix in the quark sector (5\%).

Table~\ref{tab:execosctable} summarizes the exposures needed to
achieve specific oscillation physics milestones, calculated 
for the current best-fit values of the known neutrino mixing parameters. 
Values for both the reference beam design and the optimized beamline design are shown.
For example, to reach $3\sigma$ sensitivity 
for 75\% of the range of \deltacp, a
DUNE exposure in the range of \num{850} to \SI{1320}~\ktMWyr{} is needed
for the optimized and reference beamline designs. 
Changes in the assumed value of
$\theta_{23}$ impact CP-violation and MH sensitivities the
most (discussed in \volphys) and can either reduce or increase the 
discovery potential for CP violation. To reach this level of sensitivity 
a highly capable near neutrino detector is required to control systematic uncertainties at a level lower than
the statistical uncertainties in the far detector. No experiment can provide coverage at 100\% of
\deltacp values, since CP-violating effects vanish as $\mdeltacp\to 0$
or $\pi$. Potential improvements in beamline geometry, focusing and target element designs
can significantly lower the exposure required for CP violation
discovery potential.  Several such potential improvements are discussed
in CDR \volphys and \vollbnf. 
\begin{cdrtable}[Required exposures to reach oscillation physics
  milestones]{lcc}{execosctable}{The exposure in mass (kt) $\times$ proton beam power
    (MW) $\times$ time (years) needed to reach certain oscillation physics
    milestones. The numbers are for normal hierarchy using the current best fit values of the known oscillation parameters. The two columns
    on the right are for different beam design assumptions. }
Physics milestone & Exposure \ktMWyr{} & Exposure \ktMWyr{}\\ \rowtitlestyle
  & (reference beam) & (optimized beam) \\ \toprowrule 
  $1^\circ$ $\theta_{23}$ resolution ($\theta_{23} = 42^\circ$) & 70  &  45\\ \colhline
  CPV at $3\sigma$ ($\delta_{\rm CP} = +\pi/2$)  & 70 & 60 \\ \colhline
  CPV at $3\sigma$ ($\delta_{\rm CP} = -\pi/2$)  & 160 & 100 \\ \colhline
  CPV at $5\sigma$ ($\delta_{\rm CP} = +\pi/2$)  & 280 & 210 \\ \colhline
  MH at  $5\sigma$ (worst point) & 400 & 230 \\ \colhline
  $10^\circ$ resolution ($\delta_{\rm CP} = 0$) & 450 & 290 \\ \colhline
  CPV at $5\sigma$ ($\delta_{\rm CP} = -\pi/2$)  & 525 & 320 \\ \colhline
  CPV at $5\sigma$ 50\% of \deltacp & 810 & 550 \\ \colhline 
  Reactor $\theta_{13}$ resolution & 1200 & 850 \\   
 ($\sin^2 2 \theta_{13} = 0.084 \pm 0.003$) &  &  \\ \colhline
  CPV at $3\sigma$ 75\% of \deltacp & 1320 & 850\\ 
\end{cdrtable}
  
In long-baseline experiments with \numu beams, the
magnitude of \numu disappearance and \nue appearance signals is
proportional to \sinstt{23} and \sinst{23},
respectively, in the standard three-flavor mixing scenario.  Current
\numu disappearance data are consistent with close to maximal
mixing, $\theta_{23} = 45^\circ$.  To obtain the best sensitivity to
both the magnitude of its deviation from $45^\circ$ as well the 
$\theta_{23}$ octant, a combined analysis of the two channels
is needed~\cite{Huber:2010dx}.  As demonstrated in Volume 2, a
\ktadj{40} DUNE detector with sufficient exposure will be able to
resolve the $\theta_{23}$ octant at the $3\sigma$ level or better for
$\theta_{23}$ values less than $43^\circ$ or greater than $48^\circ$.
The full LBNF/DUNE scope will allow $\theta_{23}$ to be measured with a precision of
$1^\circ$ or less, even for values within a few degrees of
$45^\circ$. 

To summarize, DUNE long-baseline program will complete
our understanding of the oscillation phenomenology. 
DUNE has great prospects to discover CP violation or, in the absence of the
effect, set stringent limits on the allowed values of \deltacp. 
DUNE will also determine the neutrino mass hierarchy with better
than a $5\sigma$~C.L.

\section{The Search for Nucleon Decay}

The DUNE far detector will significantly extend lifetime sensitivity
for specific nucleon decay modes by virtue of its high detection
efficiency relative to water Cherenkov detectors and its low
background rates.  As a LArTPC, DUNE has enhanced capability for
detecting the $p\to K^+\bar{\nu}$ channel, where lifetime
predictions from supersymmetric models extend beyond, but remain close
to, the current (preliminary) Super-Kamiokande limit of $\tau/B >
\SI{5.9e33}{year}$ (90\% C.L.), obtained from a \SI[number-unit-product = -,
inter-unit-product=\ensuremath{{}\cdot{}}]{260}{\kt\year}
exposure~\cite{kearns_isoups}\footnote{The lifetime shown here is
  divided by the branching fraction for this decay mode, $\tau/B$, and
  as such is a \emph{partial lifetime}.}.  The signature for an
isolated, nearly monochromatic charged kaon in a LArTPC is highly distinctive,
with multiple distinguishing features. 

The DUNE LArTPC far detector deep underground will reach a limit of
\SI{3e34}{\year}s after 10--12 years of operation
(Figure~\ref{fig:execsummarypdk}), depending on the deployment
scenario, and would see nine events with a background of 0.3 should
$\tau/B$ be \SI{1e34}{\year}s, just beyond the current limit. A
\ktadj{40} detector will improve the current limits by an order of
magnitude after running for two decades. Even a \ktadj{10} detector
could yield an intriguing signal of a few events after a ten-year
exposure.

\begin{cdrfigure}[Sensitivity to the decay $p\to K^+ \bar{\nu}$
  with liquid argon detectors]{execsummarypdk} {Sensitivity to the
    decay $p\to K^+ \bar{\nu}$ as a function of time for different DUNE 
LArTPC module deployment strategies. 
  For comparison, the current limit from SK is also shown, as well as the projected limit from the proposed Hyper-K experiment with \SI{5600}\ktyr{} of 
  exposure and a timeline based on a 1-Mt detector.
  The limits are at 90\% C.L., calculated for
  a Poisson process including background, assuming that the detected events
  equal the expected background.}
\includegraphics[width=0.7\textwidth]{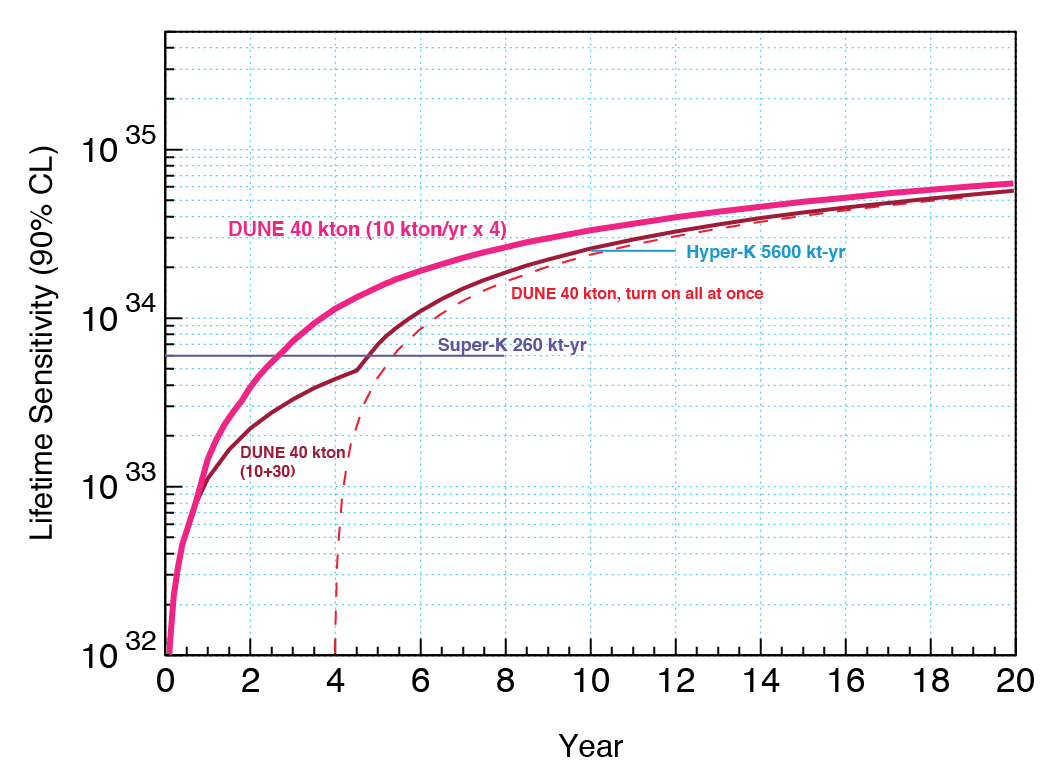}
\end{cdrfigure}

Many models in which the $p\to K^+\overline{\nu}$ channel mode is
dominant, e.g., certain supersymmetric GUT models, also favor other
modes involving kaons in the final state, thus enabling a rich program
of searches for nucleon decay in the DUNE LArTPC detector.

\section{Supernova-Neutrino Physics and Astrophysics}

The neutrinos from a core-collapse supernova are emitted in a burst of
a few tens of seconds duration, with about half the signal emitted in the first
second. The neutrino energies are mostly in the range 5--50 MeV, and the
luminosity is divided roughly equally between the three known neutrino
flavors.  Current experiments are sensitive primarily to
electron antineutrinos ($\bar{\nu}_e$), with detection through the inverse-beta decay
process on free protons\footnote{This refers to neutrino interactions with the nucleus of a
hydrogen atom in H$_2$O in water detectors or in hydrocarbon chains in 
liquid scintillator detectors.},
 which dominates the interaction rate in water
and liquid-scintillator detectors.  Liquid argon has a unique sensitivity to
the electron-neutrino ($\nu_e$) component of the flux, via the absorption
interaction on $^{40}$Ar,
\begin{eqnarray*}
\nu_e +{}^{40}{\rm Ar} & \rightarrow & e^-+{}^{40}{\rm K^*}.
\end{eqnarray*} 
This interaction can be tagged via the coincidence of the emitted
electron and the accompanying photon cascade from the $^{40}{\rm K^*}$
de-excitation.  About \num{3000} events would be expected in a \ktadj{40}
fiducial mass liquid argon detector for a supernova at a distance of
\SI{10}{\kilo\parsec}.  In the neutrino channel the oscillation
features are in general more pronounced, since the $\nu_e$ spectrum is
always significantly different from the $\nu_\mu$ ($\nu_\tau$) spectrum 
in the initial core-collapse stages, to a larger degree than is the
case for the corresponding $\bar{\nu}_e$ spectrum.  Detection of a large
neutrino signal in DUNE would help provide critical information on key
astrophysical phenomena such as
\begin{itemize}
\item the neutronization burst,
\item formation of a black hole,
\item shock wave effects,
\item shock instability oscillations, and
\item turbulence effects.
\end{itemize}

In addition to yielding unprecedented information on the mechanics of
the supernova explosion, the observation of a core-collapse supernova in
DUNE will also probe particle physics, 
providing neutrino oscillation signatures (with sensitivity to mass hierarchy and ``collective effects'' due to neutrino-neutrino interactions), as well as tests for new physics such as Goldstone bosons (e.g., Majorons), neutrino magnetic
moments, new gauge bosons (``dark photons''), ``unparticles'' and
extra-dimensional gauge bosons.

%
%
\section{Precision Measurements with the DUNE Near Detector}

The DUNE near detector
will provide precision measurements of
neutrino interactions that are essential
for controlling the systematic uncertainties in the long-baseline
oscillation physics program.  The near detector 
will include argon targets and will measure the absolute flux and energy-dependent
shape of all four neutrino species, \numu, \anumu, \nue and \anue,
to accurately predict for each species the
far/near flux ratio as a function of energy.  It will also measure the
four-momenta of secondary hadrons, such as charged and neutral mesons,
produced in the neutral- and charged-current interactions that
constitute the dominant backgrounds to the oscillation signals.

The near detector will also be the source of data for a rich program
of neutrino-interaction physics in its own right. For an integrated
beam intensity of \num{1e20} 
protons-on-target at \SI{120}{GeV}, the expected number of events per
ton is \num{170000} (\num{59000}) 
\numu (\anumu) charged-current and \num{60000} (\num{25000}) neutral-current interactions in the $\nu$ ($\overline\nu$) beam\footnote{With PIP-II, the integrated protons-on-target per year is
  expected to be around $1.1\times 10^{21}$ at \SI{120}\GeV. The mass
  of the Ar target in the DUNE ND is expected to be approximately
  100~kg.}. 
  These numbers correspond to \num{e5} neutrino interactions
on argon per year for the range of beam configurations and near detector
designs under consideration.  Measurement of fluxes, cross sections
and particle production over a large energy range of
\SIrange{0.5}{50}{\GeV} are the key elements of this program.  These
data will also help constrain backgrounds to proton-decay signals
from atmospheric neutrinos.  Furthermore, very large samples of events
will be amenable to precision reconstruction and analysis, and will be
exploited for sensitive studies of electroweak physics and nucleon
structure, as well as for searches for new physics in unexplored
regions, such as heavy sterile neutrinos, high-$\Delta m^2$
oscillations, and light Dark Matter particles. 


%
\section{Summary}


In summary, the primary science goals of DUNE are drivers for the
advancement of particle physics. The questions being addressed are of
wide-ranging consequence: the origin of flavor and the generation
structure of the fermions, 
the physical mechanism that provides the CP
violation needed to generate the baryon asymmetry of the universe, 
and the high-energy physics that would lead to the instability
of matter.  Achieving these goals requires a dedicated, ambitious and
long-term program.  No other proposed long-baseline neutrino
oscillation program with the scientific scope and sensitivity of DUNE
is as advanced in terms of engineering development and project
planning.  The staged implementation of
the far detector as four 10-kt modules will enable
exciting physics in the intermediate term, including a definitive mass
hierarchy determination and possibly a measurement of the CP phase, 
while providing the fastest route toward achieving the
full range of DUNE's science objectives.  Should DUNE find that the CP
phase is not zero or $\pi$, it will have found strong indications
($>3\sigma$) of leptonic CP violation.

The DUNE experiment is a world-leading international physics
experiment, bringing together the 
international neutrino community as well as leading experts in nucleon decay
and particle astrophysics to explore key questions at the forefront of
particle physics and astrophysics. The highly capable beam and
detectors will enable a large suite of new physics measurements with
potentially groundbreaking discoveries.

%% file: volume-project/chapter-tech-designs.tex
\chapter{Technical Overview}
\label{v1ch:tech-designs}

\section{LBNF Project}
 
To enable the scientific program of DUNE, LBNF will provide facilities that are geographically separated into the \textit{near site facilities}, those to be constructed at Fermilab, and the \textit{far site facilities}, those to be constructed at SURF.

\subsection{Near Site Facilities}
 
The scope of LBNF at Fermilab encompasses provision of the beamline plus the conventional facilities (CF) for this beamline as well as for the DUNE near detector. The layout of the near site facilities is shown in Figure~\ref{fig:nearsite-topo}. The science requirements as determined by the DUNE
collaboration drive the performance requirements of the beamline and near detector, which in turn 
dictate the requirements on  
the components, space, and functions necessary to construct, install, and operate the beamline and near detector. ES\&H and facility operations requirements (i.e., \textit{programmatic} requirements) also provide input to the design.
 
\begin{cdrfigure}[Layout of LBNF Near Site]{nearsite-topo}{Layout of the LBNF Near Site}
  \includegraphics[width=\textwidth]{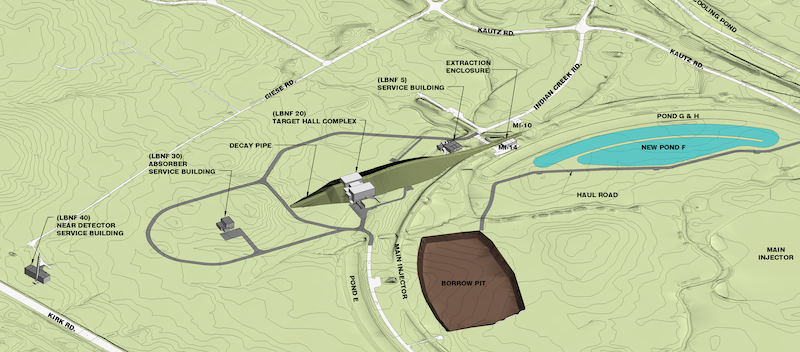}
\end{cdrfigure}

The beamline is designed to provide a neutrino beam of sufficient intensity and appropriate energy range to meet the goals of DUNE for long-baseline neutrino oscillation physics. The design is a conventional, horn-focused neutrino beamline. The components of the beamline will be designed to extract a proton beam from the Fermilab Main Injector (MI) and transport it to a target area where the collisions generate a beam of charged particles that are focused by the neutrino horns. The focused charged particles then decay, producing neutrinos (e.g., $\pi^+\rightarrow \mu^+\nu_\mu$) to create the neutrino beam directed towards the near and far detectors.
 
The facility is designed for initial operation at a proton-beam power of \SI{1.2}{\MW}, capable of supporting an upgrade to \SI{2.4}{\MW}. Operation of the facility is planned for twenty years, while the lifetime, 
including the shielding, is planned for thirty years. It is 
assumed that operations during the first five years will be at \SI{1.2}{\MW} and the remaining fifteen years at \SI{2.4}{\MW}.  
The experience gained from the various neutrino projects has contributed extensively to the reference design. In particular, the NuMI beamline serves as the prototype design. Most of the subsystem designs and the integration between them follow, to a large degree, from previous projects. 
 
The proton beam will be extracted at MI-10, a new extraction point. After extraction, this primary beam will travel horizontally heading west-northwest toward 
the far detector. It is then 
bent upward to an apex before being bent downward at the appropriate angle. 
This requires construction of an earthen embankment, or hill, whose dimensions are commensurate with the bending strength of the dipole magnets required for the beamline. 
The raised design of the primary beam 
minimizes expensive underground construction; it also significantly enhances ground-water radiological protection.

The narrow proton beam impinges on a target, producing a more diffuse, secondary beam of particles that in turn decay to produce the neutrino beam. 
The secondary pions and kaons are then focused by the neutrino horn system into a long unobstructed decay tunnel. 
The decay tunnel in the reference design is a pipe of circular cross section with its 
diameter and length optimized such that decays of the pions and kaons result in neutrinos in the energy range useful for the experiment. 
The decay tunnel is followed immediately by the absorber, which removes the remaining beam hadrons. 
 
Radiological protection is integrated into the LBNF beamline reference design in two important ways. First, shielding is optimized to reduce exposure of personnel to radiation dose and to minimize radioisotope production in ground water within the surrounding rock. Secondly, the safe handling and control of tritiated ground water produced in or near the beamline drives many aspects of the design. 
 
Beamline CF includes an enclosure connecting to the existing Main Injector at MI-10, concrete underground enclosures for the primary beam, targetry, horns and absorber, and related technical support systems. Service buildings will be constructed to provide support utilities for the primary proton beam at LBNF~5 and to support the absorber at LBNF~30 (shown in Figure~\ref{fig:nearsite-topo}).  The Target Hall Complex at LBNF~20 houses the targetry system.  Utilities will be extended from nearby existing services, including power, domestic and industrial water, sewer, and communications. 
 
Near Detector CF includes a small muon alcove area in the Beamline Absorber Hall and a separate underground Near Detector Hall that houses the near detector. A service building called LBNF~40 with two shafts to the underground supports the near detector. The underground hall is sized for the reference design near detector.
 
\subsection{Far Site Facilities}
 
The scope of LBNF at SURF includes both conventional facilities and cryogenics infrastructure to support the DUNE far detector. Figure~\ref{fig:lbnf-cavern-layout} shows the layout of the underground caverns that will house the detector modules with a separate cavern to house utilities and cryogenics systems. The requirements derive from the DUNE collaboration science requirements, which drive the space and functional requirements for constructing and operating the far detector.  ES\&H and facility operations (programmatic) requirements also provide input to the design. The far detector is divided into four \ktadj{10} fiducial mass detector modules. The designs of the four detector chambers, two each in two caverns, and the services to the caverns will be as similar to one another as possible for efficiency in design, construction and operation. 
 
\begin{cdrfigure}[LBNF Far Site cavern configuration]{lbnf-cavern-layout}{LBNF Far Site cavern configuration}  
\includegraphics[width=.7\textwidth]{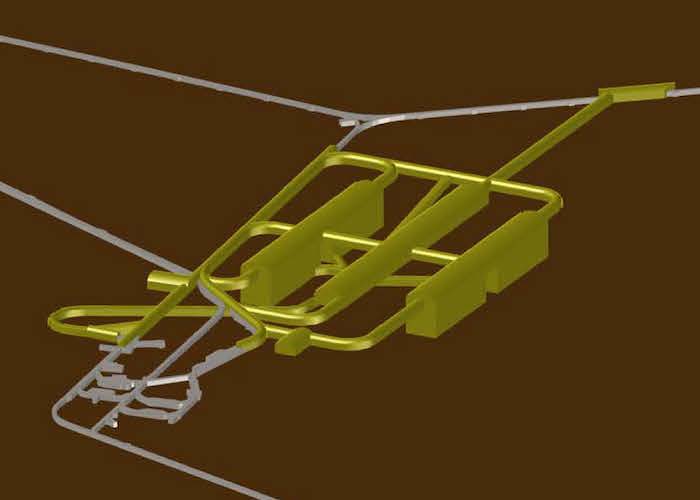}
\end{cdrfigure}

The scope of the Far Site CF includes design and construction for facilities both on the surface and underground. The underground conventional facilities include new excavated spaces at the 4850L for the detector, utility spaces for experimental equipment, utility spaces for facility equipment, drifts for access, as well as construction-required spaces. Underground infrastructure provided by CF for the detector 
includes power to experimental equipment, cooling systems and cyberinfrastructure. Underground infrastructure necessary for the facility includes domestic (potable) water, industrial water for process and fire suppression, 
fire detection and alarm, normal and standby power systems, a sump pump drainage system for native and leak water around the detector, water drainage to the facility-wide pump-discharge system, and cyberinfrastructure for communications and security.
In addition to providing new spaces and infrastructure underground, CF will enlarge and provide infrastructure in some existing spaces for LBNF and DUNE use, such as the access drifts from the Ross Shaft to the new caverns. New piping will be provided in the shaft for cryogens (gas argon transfer line and the compressor suction and discharge lines) and domestic water as well as power conduits for normal and standby power and cyberinfrastructure. 
 
SURF currently has many surface buildings and utilities, some of which will be utilized for LBNF. The scope of the above-ground CF includes only that work necessary for LBNF, and not for the general rehabilitation of buildings on the site, which remains the responsibility of SURF. 
Electrical substations and distribution will be upgraded to increase power and provide standby capability for life safety. Additional surface construction 
includes a small control room in an existing building and a new building to support cryogen transfer from the surface to the underground near the existing Ross Shaft.
 
To reduce risk of failure of essential but aging support equipment during the construction and installation period, several SURF infrastructure operations/maintenance activities are included as early activities in  the LBNF Project. These include completion of the Ross Shaft rehabilitation, rebuilding of hoist motors, and replacement of the Oro Hondo fan; if not addressed, 
failure of this aging infrastructure is more likely, which could limit or remove access to the underground areas.
 
The scope of the LBNF cryogenics infrastructure includes the design, fabrication and installation of four cryostats to contain the liquid argon (LAr) and the detector components. It also includes a comprehensive cryogenics system that meets the performance requirements for purging, cooling and filling the cryostats, for achieving and maintaining the LAr temperature, and for purifying the LAr outside the cryostats. 
 
Each cryostat is composed of a free-standing steel-framed structure with a membrane cryostat 
installed inside, to be constructed in one of the four excavated detector chambers. The cryostat is designed for a total LAr mass capacity of 
17.1~kt. Each tank has a stainless steel liner (membrane) as part of the 
system to provide full containment of the liquid. The hydrostatic pressure loading of the LAr is transmitted through rigid foam insulation to the surrounding structural steel frame, which provides external support for the cryostat. All penetrations into the interior of the cryostat will be made through the top plate to minimize the potential for leaks, with the exception of the sidewall penetration that is used for connection to the LAr recirculation system.
 
Cryogenics system components are located both on the surface and within the cavern. The cryogen receiving station is located on the surface near the Ross Shaft to allow for receipt of LAr deliveries for the initial filling period; it also has a buffer volume to accept LAr during the extended fill period. A large vaporizer for the nitrogen circuit feeds gas to one of four compressors located in the Cryogenic Compressor Building; the compressor discharges high-pressure nitrogen gas to pipes in the Ross shaft. The compressors are located on the surface because the electrical power  and thermal cooling requirements are less stringent than for an installation at the 4850L.  
 
Equipment at the 4850L includes the nitrogen refrigerator, liquid nitrogen vessels, argon condensers, external LAr recirculation pumps, and filtration equipment. Filling each cryostat with LAr in a reasonable period of time is a driving factor for the refrigerator and condenser sizing.  Each cryostat will have its own argon recondensers, argon-purifying equipment and overpressure-protection system located in the Central Utility Cavern. Recirculation pumps will be placed outside of and adjacent to each cryostat in order to circulate liquid from the bottom of the tank through the purifier.
 
\section{Strategy for Developing the LBNF Beamline}
\label{v1ch:tech-designs:beam-strategy}

The neutrino beamline described in this CDR is a direct outgrowth of the design~\cite{lbnecdr} developed for the LBNE
CD-1 review in 2012.  That design was driven by the need to minimize cost, while delivering the performance required to meet the scientific objectives of the long-baseline neutrino program.  It includes many features that followed directly from the 
successful NuMI beamline design as updated for the NOvA experiment.  It utilizes a target and horn system based on NuMI designs, with the spacing of the target and two horns set to maximize flux at the first, and to the extent possible, second 
oscillation maxima, subject to the limitations of  the NuMI designs for these systems.  The target chase volume --- length and width --- are set to the minimum necessary to accommodate this focusing system, and the temporary morgue space to store 
used targets and horns is sized based on the size of the NuMI components.  Following the NuMI design, the decay pipe is helium-filled, while the target chase is air-filled.  
 
The LBNF beamline is designed to utilize the Main Injector proton beam, as will be delivered after the PIP-II upgrades~\cite{pip2-2013}.  
The  proton beam energy can be chosen to be between 60 and 120~GeV, with the corresponding range of beam power from 1.0 to 1.2~MW.
The ability to vary the 
proton beam energy is important for optimizing the flux spectrum and to understand systematic effects in the beam production, and to provide flexibility to allow the facility to address future questions in neutrino physics that may require a 
different neutrino energy spectrum.  To allow for the higher beam power that will be enabled by future upgrades to the Fermilab accelerator complex beyond PIP-II, the elements of the beamline and supporting conventional facilities that cannot be changed once 
the facility is built and has been irradiated are designed to accommodate beam power in the range of 2.0 to 2.4~MW for the corresponding proton beam energy range of 60 to 120~GeV.  These elements include primary beam components, target hall, decay pipe and absorber, as 
well as the shielding for them.  Components that can be replaced, such as targets and horns, are designed for the \MWadj{1.2} initial operation.  
Additional R\&D
will be required to develop these components 
for operation at the higher beam power. 

Since the 2012 CD-1 review, the beamline design has evolved in a number of areas, as better understanding of the design requirements and constraints has developed.  Some of these design changes have come to full maturity and are 
described in this CDR.  Others require further development and evaluation to determine if and how they might be incorporated into the LBNF neutrino beamline design.  They offer the possibility of higher performance, flexibility in 
implementation of future ideas, and/or greater reliability and will be developed by the DUNE
collaboration in the near future. The beamline facility is designed to have an operational lifetime of 20 years, and it is important that it be designed to allow future upgrades and modifications that will allow it to 
exploit new 
technologies and/or adapt the neutrino spectrum to address new questions in neutrino physics over this long period. The key alternatives and options under consideration and the strategy for evaluating and potentially implementing them are summarized below.  They are described in more detail in the \href{https://web.fnal.gov/project/LBNF/SitePages/Proposals%20and%20Design%20Reports.aspx}{other volumes of this CDR and in its annexes}.

Further optimization of the target-horn system has the potential to substantially increase the neutrino flux at the first and especially second oscillation maxima and to reduce wrong-sign neutrino background, thereby increasing the sensitivity to CP 
violation and mass hierarchy determination, as discussed in \volphys.  This optimization work is ongoing and may yield further improvements beyond those currently achieved. Engineering studies of the proposed horn designs and methods 
of integrating the target into the first horn must be performed to turn these concepts into real structures that can be built and that satisfy additional requirements in areas such as reliability and longevity.  These studies will be carried out between CD-1 and CD-2 to 
determine the baseline design for the LBNF target-horn system.  Since targets and horns must be replaceable, it is also possible to continue development of the target-horn system in the future and replace the initial system with a more advanced 
design or one optimized for different physics.  Such future development, beyond that necessary to establish the baseline design at CD-2, would be done outside of the LBNF Project.
 
The more advanced focusing system, called the ``optimized beam configuration'' in \volphys, utilizes horns that are longer and larger in diameter and that are spaced farther apart than in the reference design, which would require a target chase approximately 9~m longer and 0.6~m wider.  
It cannot be ruled out that further optimization, or future designs that would
allow exploration of new questions may require additional space beyond this.  Also, the larger horns will require a larger 
space for temporary storage of used, irradiated components, requiring, in turn, an increase in the size of the morgue or a revision of the remote handling approach.  Between CD-1 and CD-2, studies will be done to determine not only the geometric 
requirements from the final baseline target-horn system, but also to estimate the dimensions needed to accommodate potential future designs.
 
The material, geometry and structure of the target assembly itself can have significant impact both on the effective pion production and the energy spectrum of pions, which in turn affect the neutrino spectrum, and on the reliability and longevity of 
the target, which affects the integrated beam exposure.  Potential design developments range from incremental (e.g., changing from the reference design rectangular cross section, water-cooled graphite target to a cylindrical 
helium-cooled target), to more substantial (e.g., changing target material from graphite to beryllium), to radical (e.g., implementing a hybrid target with lighter material upstream and heavier material downstream and perhaps constructed of a set of spheres captured in a 
cylindrical skin).  New designs beyond the current reference design are also needed in order to accommodate the higher beam power (up to 2.4~MW) that will be provided by the PIP-III upgrade.  
Target development will largely be carried out in the 
context of worldwide collaborations on high-power targetry such as the Radiation Damage In Accelerator Target 
Environments (RaDIATE~\cite{radiate-web}) collaboration, and not within the LBNF Project. The LBNF design must be such that it can fully exploit future developments in target design.
 
The length and diameter of the decay pipe also affect the neutrino flux spectrum.  A longer decay pipe increases the total neutrino flux with a larger increase at higher energies; a larger diameter allows the capture and decay of lower-energy pions, 
increasing the neutrino flux at lower energies as described in \volphys. The dimensions also affect the electron-neutrino and wrong-sign backgrounds.  Unlike targets and horns, the decay pipe cannot be modified after the facility is built, making the 
choice of geometry particularly important.  The reference design values of \SI{204}{\meter} length and \SI{4}{\meter} diameter appear well matched to the physics of DUNE but studies to determine the optimal dimensions continue.  The cost of increasing the decay 
pipe length or diameter  is relatively large, including 
the impact on the absorber.
Therefore, studies of the decay pipe must include 
evaluation of the relative advantages of
investment in the decay pipe versus investment in 
other systems, e.g., a larger target hall complex, more advanced target-horn systems, or more far detector mass.  
Studies currently in progress will continue to be carried out jointly by LBNF and DUNE between CD-1 and CD-2 to determine the baseline decay-pipe geometry.

\section{DUNE Detectors}

The DUNE detectors to be installed at SURF (the far site) and Fermilab (the near site) will enable the scientific program of DUNE.  The detector 
requirements derive from the DUNE science goals.

\subsection{The Far Detector}
The  far detector will be located deep underground at the 4850L and have
a  fiducial mass of \SI{40}{\kt} to perform sensitive studies of long-baseline oscillations with a \kmadj{1300} baseline as well as a rich astroparticle physics program and nucleon decay searches. The far detector  will be composed of four 
similar modules, each instrumented as a liquid argon time-projection chamber (LArTPC).
The concept of the LArTPC provides
excellent tracking and calorimetry performance, hence it is ideal for massive neutrino detectors such as the DUNE far detector, which require high signal efficiency and effective background discrimination,  an excellent capability to identify and  precisely measure neutrino events over a wide range of energies, and an excellent reconstruction of the kinematical properties
with a high resolution. The full imaging of events will allow study of neutrino interactions and
other rare events with an unprecedented resolution.
 The huge mass will allow collection of sufficient statistics for precision
studies, as discussed in Chapter~\ref{v1ch:science}.

The LArTPC, pioneered in the context of the ICARUS project, is a mature technology. It is the outcome
of several decades of worldwide R\&D.  Nonetheless, the size of a single \ktadj{10} DUNE module represents an extrapolation of over one order of magnitude compared to the largest operated detector, the ICARUS~T600. To address this challenge, DUNE is developing two far detector options, the reference design and an alternative design, and is engaged in a 
comprehensive prototyping effort. At this stage, the development of two options is a strength 
made possible by the merging of the worldwide neutrino community into DUNE.  The two detector
concepts are illustrated in Figure~\ref{fig:FarDet-overview-SPDP}.

\begin{cdrfigure}[3D models of the DUNE far detector designs]{FarDet-overview-SPDP}
{3D models of two 10-kt detectors using the single-phase reference design (left) 
and the dual-phase alternative design (right) for the DUNE far detector to be 
located at 4850L.}
\centering
\begin{minipage}[b]{1.0\textwidth}
\begin{center}
\includegraphics[width=.5\textwidth]{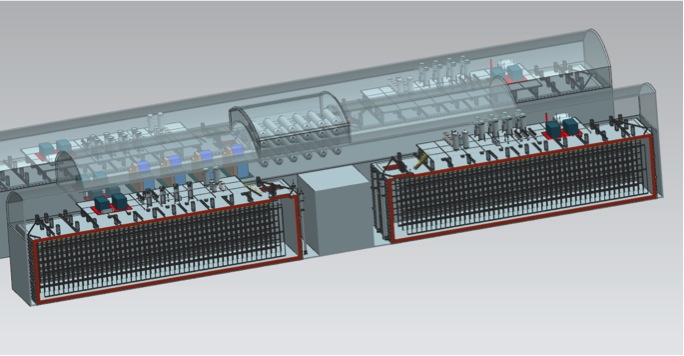}
\includegraphics[width=0.46\textwidth]{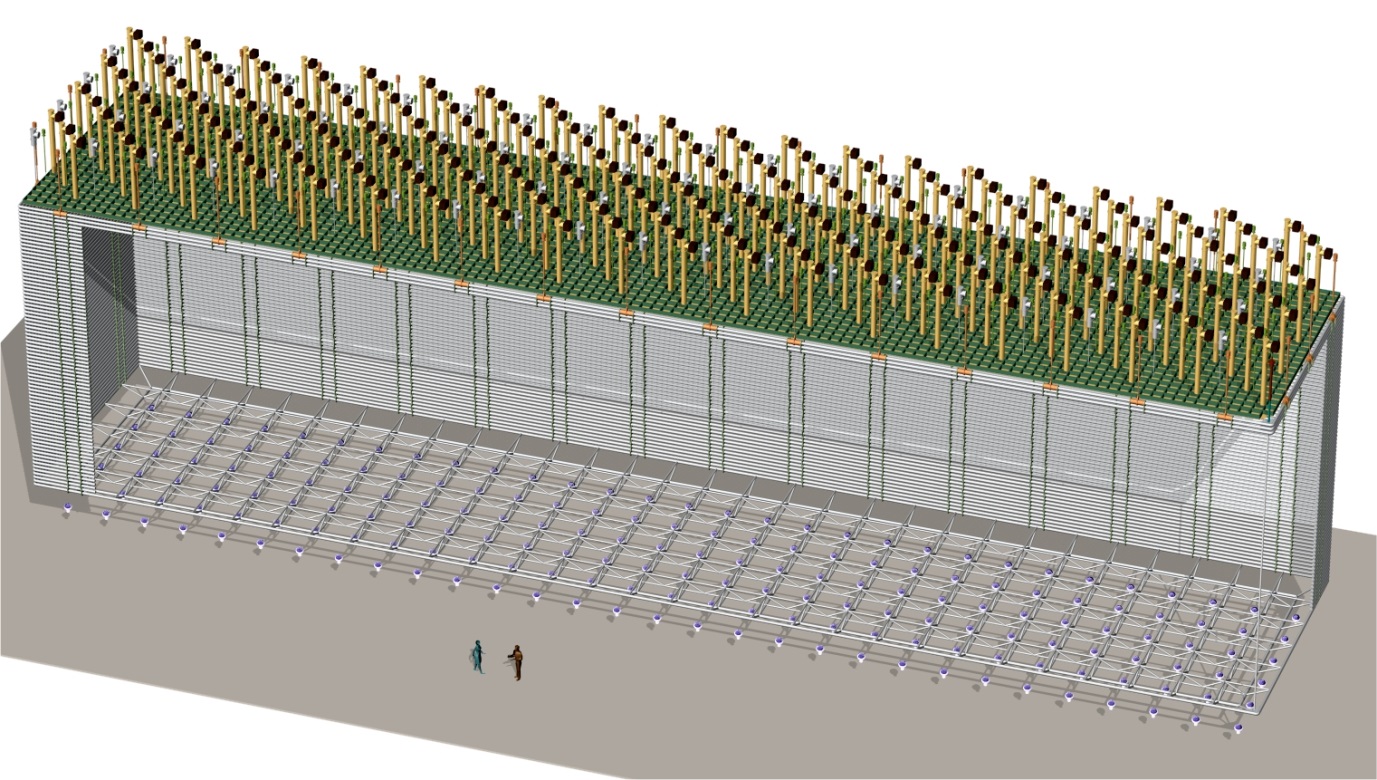}
\end{center}
\end{minipage}
\end{cdrfigure}

Interactions in LAr produce ionization charge and scintillation light.
The ionization electrons are drifted with a constant electric field away from the cathode
plane and towards the segmented anode plane. 
The prompt scintillation light,
detected by photo-detectors, provides the absolute time of the event.
The reference design adopts a single-phase readout, where the readout anode is composed of wire planes in the LAr volume. 
The alternative design implements a  dual-phase approach, in which the 
ionization charges are extracted, amplified and detected in gaseous argon (GAr) above the liquid surface. 
The dual-phase design would allow for a finer readout pitch (3~mm), 
a lower detection-energy threshold, and better pattern reconstruction of the events.
The photon-detection schemes used in the two designs are complementary, one is distributed
within the LAr volume, the other is concentrated at the bottom of the tank.

The \ktadj{10} reference design TPC is described in Chapter 4 of \voldune. 
Its active volume is 12\,m high, 14.5\,m wide and 
58\,m long, instrumented with anode plane assemblies (APAs), 
which are 6.3\,m high and 2.3\,m wide, and cathode plane assemblies (CPAs), 3\,m high by 2.3\,wide. 
Vertical stacks of
two APAs and four CPAs 
instrument the 12\,m height of the active volume. The 12.5-m width of the detector is 
spanned by three stacks of APAs and two stacks of CPAs in an APA:CPA:APA:CPA:APA
arrangement, resulting in four 3.6-m drift volumes, while the 58-m length of the active volume
is spanned by 25 such stack arrangements placed edge to edge. Hence a \ktadj{10} 
far detector module consists of 150 APAs and 200 CPAs. The CPAs are held at $-$180\,kV, such that 
ionization electrons drift a maximum distance of 3.6\,m in the electric field of 500\,V\,cm$^{-1}$.
The highly modular nature of the detector design allows for manufacturing to be distributed across a number of sites.

A  comprehensive prototyping strategy for both designs is actively pursued (see Chapter 9 of \voldune).
The reference design, closer to the original ICARUS design, is currently being validated in the 35-t prototype 
LAr detector at Fermilab.  The alternative design, representing a novel approach, has been proven on several
small-scale prototypes. Presently
a 20-t dual-phase prototype (WA105) with dimensions 3$\times$1$\times$1~m$^3$ is being constructed at CERN,  
and should be operational in 2016. 
The ultimate validation of the engineered solutions for both designs of the FD is foreseen in
the context of the neutrino activities at the CERN North Area extension (EHN1 area) around 2018, 
where full-scale engineering prototypes will be 
assembled and commissioned. Following this milestone, a test-beam data 
campaign will be executed 
to collect a large sample of charged-particle interactions
in order to study the response of the detector with high precision.
A comprehensive list of synergies between the reference and alternative designs has been identified (Chapter 6 of \voldune). Common solutions for DAQ, electronics, HV feed-throughs, and so on, will pursued and implemented, independent of the details of the TPC design. The ongoing and planned efforts 
will
provide the ideal environment to exploit such synergies and implement common solutions.
There is recognition that the LArTPC technology will continue to evolve with (1) the large-scale prototypes at the CERN Neutrino Platform and the experience from the Fermilab SBN program, and (2) the experience gained during the construction and commissioning of the first \ktadj{10} module. 
The staged approach with the deployment of consecutive modules will
enable an early science program while allowing implementation of improvements and developments  during the experiment's lifetime.
The strategy for implementing
the far detector is presented in Section~\ref{v1ch3:fd-impl-strategy}.

\subsection{The Near Detector} 

The primary role of the DUNE near detector system is to characterize the energy spectrum and the composition of the neutrino beam at the source, in terms of
both muon- and electron-flavored neutrinos and antineutrinos, and to provide measurements of neutrino interaction cross sections. This is 
necessary to control systematic uncertainties with the precision needed to fulfill the DUNE primary science objectives.
The separation between fluxes of neutrinos and antineutrinos requires a magnetized neutrino detector to 
charge-discriminate electrons and muons produced in the neutrino charged-current interactions.
 As the near detector will be exposed to an intense flux of neutrinos, it will 
collect an unprecedentedly large sample of neutrino 
interactions, allowing for an extended science program. 
The near detector will therefore provide a broad program of fundamental neutrino interaction 
measurements, which are an important part of the ancilliary scientific goals of the DUNE collaboration. 
The reference design for the near detector design is the NOMAD-inspired fine-grained tracker (FGT), illustrated in Figure~\ref{fig:FGT_schematic}. Its subsystems include a central 
straw-tube tracker and an electromagnetic calorimeter embedded in a 0.4-T dipole field. The steel of the
magnet yoke will be instrumented with muon identifiers. The strategy for implementation of
the near detector is presented in Section~\ref{v1ch:strategyND}.

\begin{cdrfigure}[Fine-grained
tracker (FGT) near neutrino detector schematic]{FGT_schematic}{A schematic drawing of the fine-grained tracker design}
\includegraphics[width=.8\textwidth]{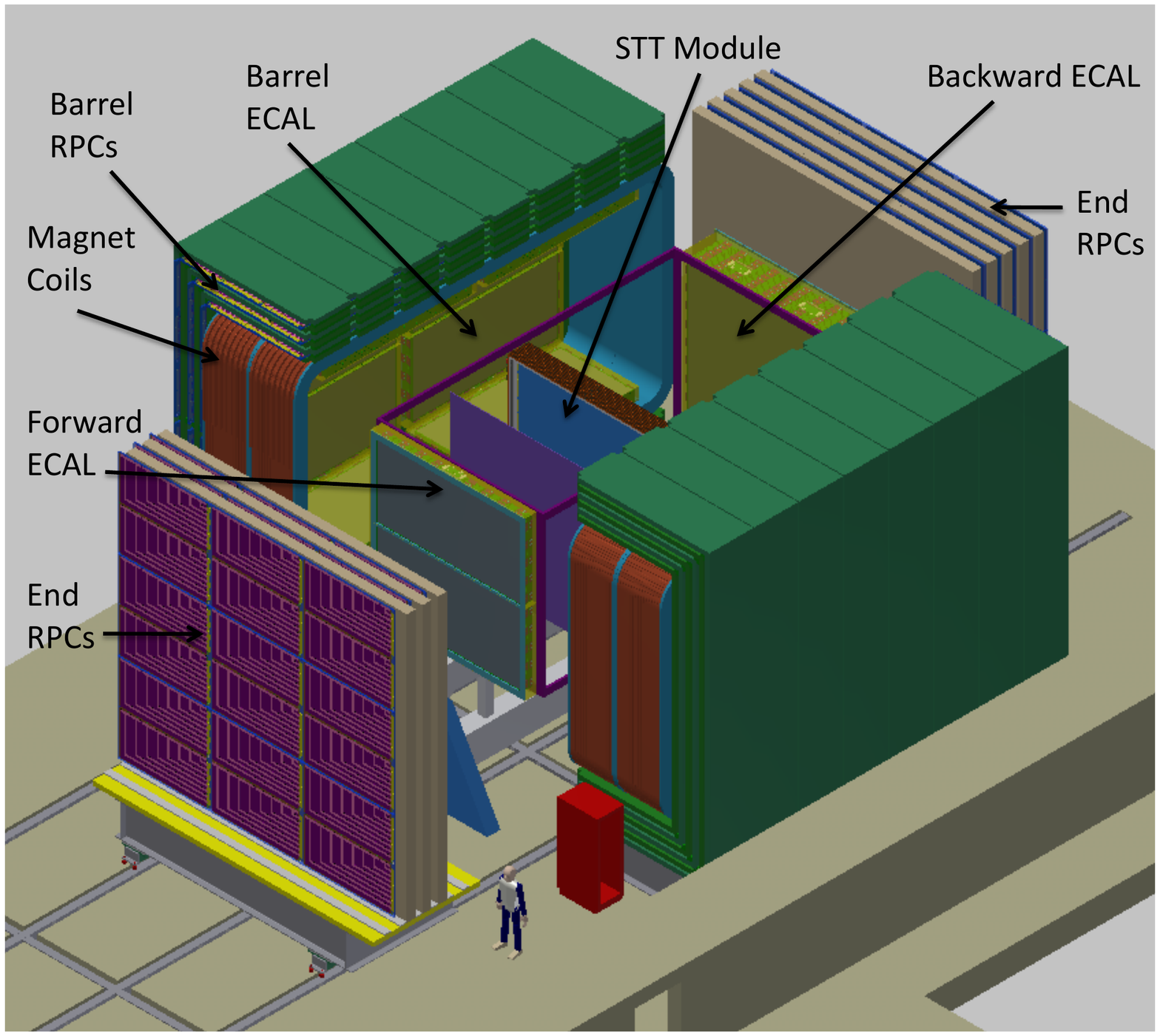}
\end{cdrfigure}

The near detector will be complemented by a Beamline Measurement System (BLM) located in the region of the beam absorber at the downstream end of the decay region. The BLM aims to measure the muon fluxes from hadron decay and
 is intended to monitor the beam profile on a spill-by-spill basis. It will operate for the life of the experiment.

\section{Strategy for Implementing the DUNE Far Detector}
\label{v1ch3:fd-impl-strategy}

The LBNF project will provide four separate cryostats to house the far detector (FD) modules on the 4850L at SURF.  
Instrumentation of the first detector module
will commence in 2021. 
As part of the deployment and risk mitigation strategies, 
the cryostat for the second detector must be available when the first cryostat 
is filled. The aim is to install the third and fourth cryostats as rapidly thereafter as funding 
allows.

The DUNE collaboration aims to deploy four 10-kt (fiducial) mass FD modules based 
on the 
LArTPC technology, the viability 
of 
which has been proven by the ICARUS experiment. Neutrino 
interactions in liquid argon produce ionization and scintillation signals. While 
the basic detection method is the same, DUNE contemplates two options for the readout 
of the ionization signals: single-phase readout, where the ionization is detected 
using readout (wire) planes in the liquid argon volume; and the dual-phase approach, where 
the ionization signals are amplified and detected in gaseous argon above the liquid 
surface. The dual-phase approach, if demonstrated, would allow for a 3-mm readout 
pitch, a lower detection energy threshold, and better reconstruction of 
the events. 
An active development program for 
both technologies is being pursued in the context of the Fermilab Short-Baseline Neutrino (SBN)
program and 
the CERN Neutrino Platform. 
A flexible 
approach to the DUNE far detector designs offers the potential to bring additional 
interest and resources into the experimental collaboration. 

\subsection{Guiding Principles for the DUNE Far Detector}

\begin{itemize}
\item The lowest-risk design for the first 10-kt module satisfying the requirements 
will be adopted, allowing for its installation at SURF to commence in 2021. 
Installation  of the second 10-kt module should commence before 2022. 

\item  There is recognition that the LArTPC technology will continue to evolve with: (1) the 
large-scale prototypes at the CERN Neutrino Platform and the experience from the 
Fermilab SBN program, and (2) the experience gained during the construction and 
commissioning of the first 10-kt module. It is assumed that all four modules 
will be similar but not necessarily identical.

\item  In order to start installation on the timescale of 2021, 
the first  10-kt module will be based on the APA/CPA (single-phase) design, which is currently the lowest
risk option. There will be a clear and transparent decision process (organized by the DUNE 
Technical Board) for the design 
of the second and subsequent far detector modules, allowing for evolution of the 
LArTPC technology to be implemented. The decision will be 
based on physics performance, technical and schedule risks, costs and funding 
opportunities.

\item The DUNE collaboration will instrument the second cryostat as soon as possible.

\item A comprehensive list of synergies between the reference and alternative designs 
has been identified and summarized in \voldune. Common solutions for DAQ, 
electronics, HV feed-throughs, etc., will be pursued and implemented, independent 
of the details of the TPC design.

\end{itemize}
\subsection{Strategy for the First 10-kt Far Detector Module}
\label{v1ch3:first-fd-mod-strategy}

The viability of wire-plane LArTPC readout has already been demonstrated by the ICARUS T600 
experiment, where data were successfully accumulated over a period of three years. 
An extrapolation of the observed performance and the implementation of improvements 
in the design (such as immersed cold electronics) will allow the single-phase 
approach to meet the LBNF/DUNE far detector requirements. In order to start the FD installation
by 2021, 
the first 10-kt module will be based on the single-phase design using anode and cathode
plane assemblies (APAs and CPAs), described in Chapter 4 of \voldune. 
Based on previous experience and the 
future development path in the Fermilab SBN program and at the CERN Neutrino Platform, 
this choice represents the lowest-risk option for installation of the first 10-kt FD module by 
2021. 
For these reasons, the APA/CPA single-phase wire plane LArTPC readout 
concept 
is the \textit{reference design} 
for the far detector. 
The design is already relatively advanced for the conceptual  
stage. From this point on, modifications to the reference design will require approval
by the DUNE Technical Board. A preliminary design review will take place as early 
as possible, utilizing the experience from the DUNE 35-t prototype; the design 
review will define the baseline design that will form the basis of the TDR (CD-2). 
At that point, the design will be put under a formal change-control 
process. 

A single-phase engineering prototype,
comprising six full-sized drift cells of the TDR engineering baseline,
is planned as a central part of the risk-mitigation 
strategy for the first 10-kt module. It
will be validated at the CERN Neutrino Platform in 2018 (pending approval by CERN). 
%
Based on the  performance of this prototype, 
a final design review will take place towards the end of 2018 and construction of the readout planes will 
commence in 2019, to be ready for first installation in 2021. 
The design reviews will be organized by the DUNE Technical Coordinator. 

In parallel with preparation for construction of the first 10-kt far detector module, 
the DUNE collaboration recognizes the potential of the dual-phase technology and 
strongly endorses the already approved development program at the CERN Neutrino 
Platform (the WA105 experiment), which includes the operation of the 20-t prototype 
in 2016 and the 6$\times$6$\times$6\,m\textsuperscript{3} demonstrator in 2018. Participation 
in the WA105 experiment is open to all DUNE collaborators. A concept for the dual-phase 
implementation of a far detector module is presented as an \textit{alternative 
design} in \voldune. This alternative design, if demonstrated, 
could form the basis of the second or subsequent 10-kt modules, in 
particular to achieve improved detector performance in a cost-effective way. 

\subsection{DUNE at the CERN Neutrino Platform}
\label{v1ch3:dune-at-cern}

Two large LArTPC prototypes are in progress at the CERN Neutrino Platform, a single-phase and a dual-phase, to be ready on similar time scales. 
For the dual-phase prototype, WA105 has signed an MoU with the CERN Neutrino Platform to provide the large 
$\sim$8$\times$8$\times$8\,m\textsuperscript{3} cryostat by October 2016 in the new EHN1 extension. 
Both prototypes will be exposed to a charged-particle test 
beam spanning a range of particle types and energies.  

The DUNE collaboration will instrument 
the single-phase LArTPC with an arrangement 
of six APAs and six CPAs, in an APA:CPA:APA configuration, providing an engineering 
test of the full-size drift volume. These assemblies will be produced at two or more sites with the cost 
shared between the DOE project and international partners. This CERN prototype thus 
provides the opportunity for the production sites to validate the manufacturing 
procedure ahead of large-scale production for the far detector. Three major operational 
milestones are defined for this single-phase prototype: (1) engineering validation 
(successful cool-down); (2) operational validation (successful TPC readout with 
cosmic-ray muons); and (3) physics validation with test-beam data. Reaching milestone 
2, scheduled for early 2018, will allow the retirement of a number of technical 
risks for the construction of the first 10-kt module. The proposal for the DUNE 
single-phase prototype will be presented to the CERN SPS Scientific Committee in June 2015. 

In parallel, the WA105 experiment, approved by the CERN Research Board in 2014 and supported 
by the CERN Neutrino Platform, has a funded plan to construct and operate a large-scale 
demonstrator utilizing the dual-phase readout in the test beam by October 2017. 
Successful operation and demonstration of long-term stability of the WA105 demonstrator 
will establish this technological solution as an option for the second or subsequent 
far detector modules. The DUNE dual-phase design is based on independent 3$\times$3\,m$^2$
charge readout planes (CRP) placed at the gas-liquid interface. Each module provides 
two perpendicular ``collection'' views with 3-mm readout pitch. A \ktadj{10} module 
would be composed of 80 CRPs hanging from the top of the cryostat, decoupled from 
the field cage and cathode. The WA105 demonstrator will contain four 3$\times$3m$^2$ 
CRPs, 
providing the opportunity to validate the manufacturing procedure 
ahead of large-scale production. WA105 is presently constructing a 3$\times$1\,m$^2$ 
CRP to be operated in 2016. The same operational milestones (engineering, operational, 
physics) are defined for the dual-phase as for the single-phase prototype.

The DUNE program at the CERN Neutrino Platform will be coordinated by a single 
L2 manager. Common technical solutions will be adopted wherever possible for the 
two prototypes. 
The charged-particle test-beam data will provide essential calibration samples 
for both technologies and will enable a direct comparison of the relative physics 
capabilities of the single-phase and dual-phase TPC readout. 

\subsection{Strategy for the Second and Subsequent 10-kt Far Detector Modules}
\label{v1ch3:subsqt-fd-mod-strategy}

For the purposes of cost and schedule, the reference design for the first module 
is taken as the reference design for the subsequent three modules. However, 
the experience with the first \ktadj{10} module and the development activities at 
the CERN Neutrino Platform are likely to lead to the evolution of the TPC technology, both 
in terms of refinements to single-phase design and the validation of the operation 
of the dual-phase design. The DUNE technical board will instigate a formal review 
of the design for the second module in 2020; the technology choice 
will be based on risk, cost (including the potential benefits of additional 
non-DOE funding) and physics performance (as established in the CERN charged-particle 
test beam). After the decision, the design of the second module will come under formal 
change control. This process will be repeated for the third and fourth modules. 

This strategy allows flexibility with respect to international contributions, enabling the DUNE collaboration to
adopt evolving approaches for subsequent modules. This approach provides the possibility of attracting interest 
and resources from a broader community, and space for flexibility to respond to 
the funding constraints from different sources. 

\section{Strategy for Implementing the DUNE Near Detector(s)}
\label{v1ch:strategyND}

 The primary scientific motivation for 
the DUNE near detector 
is to determine the beam spectrum for the long-baseline 
neutrino oscillation studies. The near detector, which is exposed to an intense 
flux of neutrinos, also enables a wealth of fundamental neutrino 
interaction measurements, which are an important part of the  scientific 
goals of the DUNE collaboration. Within the former LBNE collaboration the neutrino 
near detector design was the NOMAD-inspired fine-grained tracker (FGT), which 
was established through a strong collaboration of U.S. and Indian institutes.

\subsection{Guiding Principles for the DUNE Near Detector}

It is recognized that a detailed cost-benefit study of potential near detector options 
has yet to take place and such a study is of high priority to the DUNE collaboration. The
primary design considerations for the DUNE near neutrino detector include
\begin{itemize}
\item  
the 
ability to adequately constrain the systematic errors in the DUNE LBL oscillation 
analysis, which requires the capability to precisely measure exclusive neutrino
interactions; and

\item 
the self-contained non-oscillation 
neutrino physics program.

\end{itemize}

\subsection{DUNE Near Detector Reference Design }

The NOMAD-inspired fine-grained tracker (FGT) concept is the \textit{reference 
design} for CD-1 review. The cost and resource-loaded schedule for CD-1 review 
will be based on this design, as will the near site conventional facilities. The 
Fine-Grained Tracker consists of:  central straw-tube tracker (STT) of volume 
3.5\,m$\times$3.5\,m$\times$6.4\,m; a lead-scintillator sandwich sampling electromagnetic calorimeter 
(ECAL); a large-bore warm dipole magnet, with inner dimensions of 
4.5\,m$\times$4.5\,m$\times$8.0\,m, surrounding the STT and ECAL and providing a magnetic field of 0.4\,T; 
and RPC-based muon detectors (MuIDs) located in the steel of the magnet, as well 
as upstream and downstream of the STT. The reference 
design is presented in Chapter 
7 of \voldune. 

For ten years of operation in the LBNF 1.2-MW beam (5 years neutrinos + 5 years 
antineutrinos), the near detector will record a sample of more than 100 million 
neutrino interactions and 50 million antineutrino interactions. These vast samples 
of neutrino interactions will provide the necessary strong constraints on the 
systematic uncertainties for the LBL oscillation physics --- the justification is 
given in Section 6.1.1 of \volphys. The large samples of neutrino 
interactions will also provide significant physics opportunities, including 
numerous topics for PhD theses.  


The contribution of Indian institutions to the design and construction of the DUNE 
FGT neutrino near detector is a vital part of the strategy for the construction 
of the experiment. The reference design will provide a rich self-contained physics 
program. From the perspective of an ultimate LBL oscillation program, there may 
be benefits of augmenting the FGT with, for example, a relatively small LArTPC 
in front of the FGT that would allow for a direct comparison with the far detector. 
A second line of study would be to augment the straw-tube tracker  with 
a high-pressure gaseous argon TPC. At this stage, the benefits of such options 
have not been studied; alternative designs for the near detector are not presented in 
the CDR and will be the subject of detailed studies in the coming months. 

\subsection{DUNE Near Detector Task Force}

A full end-to-end study of the impact of the near detector design (in particular of the fine-grain tracker) on the LBL oscillation 
systematics has yet to be performed. Many of the elements of such a study are in 
development, for example the Monte Carlo simulation of the FGT and the adaptation 
of the T2K framework for implementing ND measurements as constraints in the propagation 
of systematic uncertainties to the far detector. 

After the CD-1-R review, the DUNE collaboration will initiate a detailed study 
of the optimization of the near detector. To this end a new task force will be set 
up with the charge of:

\begin{itemize}
\item delivering the simulation of the near detector reference design and possible alternatives,

\item undertaking an end-to-end study to provide a quantitative understanding of 
the power of the near detector designs to constrain the systematic uncertainties on the LBL 
oscillation measurements, and

\item quantifying the benefits of augmenting the reference design with a LArTPC 
or a high-pressure gaseous argon TPC.
\end{itemize}

High priority will be placed on this work and the intention is to engage a broad 
cross section of the collaboration in this process. The task force will be charged 
to deliver a first report by July 2016. Based on the final report of this task force and input 
from the DUNE Technical Board, the DUNE Executive Board will refine the DUNE strategy 
for the near detector.

%% file: volume-project/chapter-org-mgmt.tex
\chapter{Organization and Management}
\label{v1ch:org-mgmt}

\section{Overview}

To accommodate a variety of international funding model constraints, LBNF and DUNE are organized as separate projects. As mentioned in the Introduction, the LBNF project is responsible for design and construction of the conventional facilities, beamlines, and cryogenic infrastructure needed to support the experiment.  The DUNE project is responsible for the construction and commissioning of the detectors used to pursue the scientific program.  LBNF is organized as a DOE/Fermilab project incorporating international partners.   DUNE is an international project organized by the DUNE collaboration with appropriate oversight from stakeholders including the DOE.

\section{LBNF}

\subsection{Project Structure and Responsibilities}

The LBNF project is charged by Fermilab and DOE to design and construct the conventional and technical facilities needed to support the DUNE collaboration.  LBNF works in close coordination with DUNE to ensure that the scientific requirements of the program are satisfied through the mechanisms described in Section~\ref{sec:lbnf-dune-interface}. LBNF also works closely with SURF management to coordinate the design and construction of the underground facilities required for the DUNE far detector. 

LBNF consists of two major L2 subprojects coordinated through a central Project Office located at Fermilab: Far Site Facilities and Near Site Facilities. Each L2 project incorporates several large L3 subprojects as detailed in the WBS structure presented in Figure~\ref{fig:lbnf-wbs}.

The project team consists of members from Fermilab, CERN, South Dakota Science and Technology Authority (SDSTA), and BNL.  The team, including members of the Project Office as well as the L2 and L3 managers for the individual subprojects, is assembled by the Project Director. The project team to WBS Level~3 is shown in Figure~\ref{fig:lbnf-org}. 
Line management for environment, safety and health, and quality assurance flows through the Project Director.

Through their delegated authority and in consultation with major stakeholders, the L2 Project Managers determine which of their lower-tier managers will be Control Account Managers (CAMs) for the project WBS. L2 and L3 Project Managers are directly responsible for generating and maintaining the cost estimate, schedule, and resource requirements for their subprojects and for meeting the goals of their subprojects within the accepted baseline cost and schedule. 

\begin{cdrfigure}[LBNF Work Breakdown Structure (WBS) to level 3]{lbnf-wbs}{LBNF Work Breakdown Structure (WBS) to level 3}
  \includegraphics[width=0.8\textwidth]{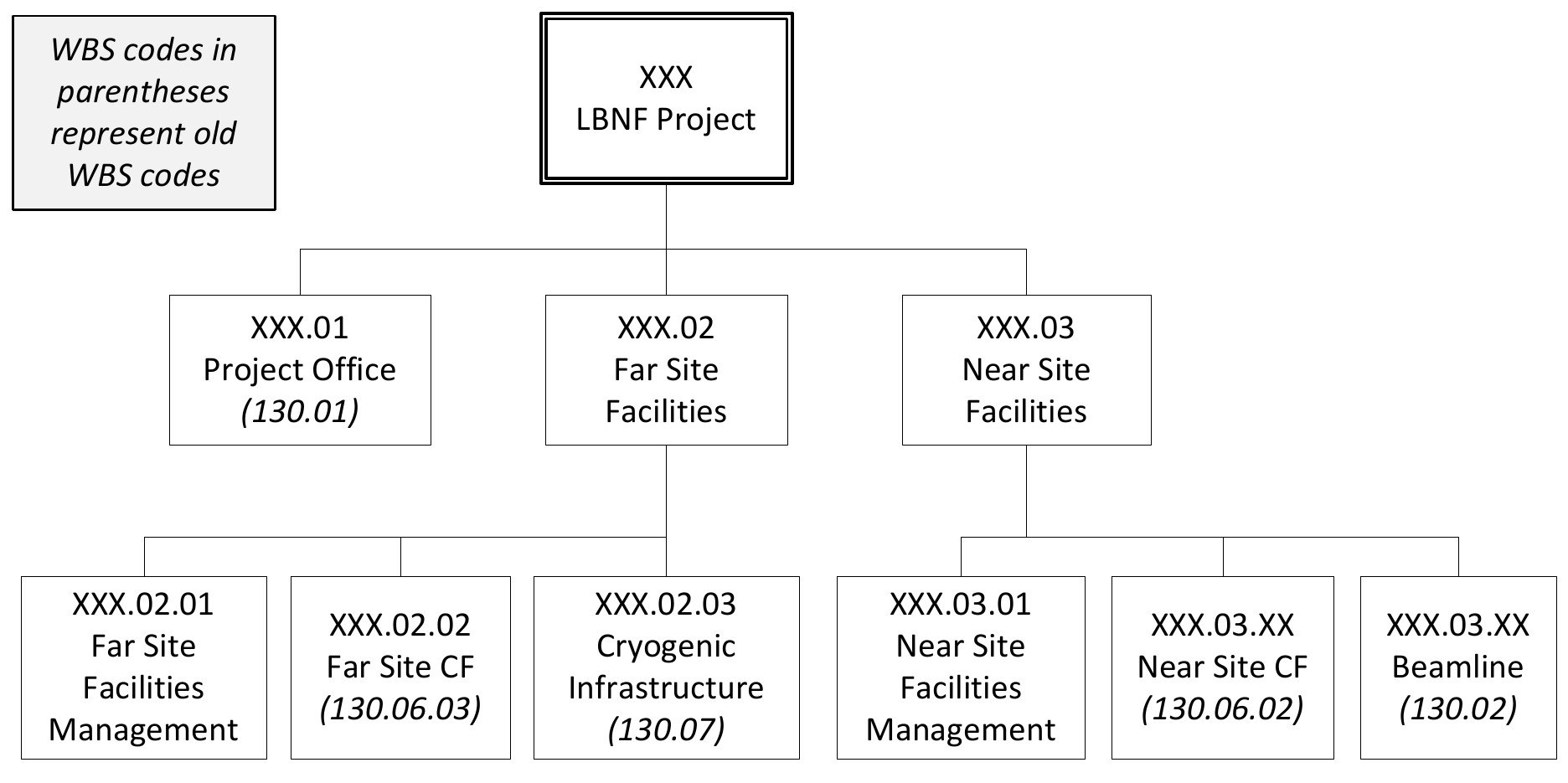}
\end{cdrfigure}

The design and construction of LBNF is supported by other laboratories and consultants/contractors that provide scientific, engineering, and technical expertise. A full description of LBNF Project Management is contained within the LBNF Project Management Plan\cite{PMP-10770}. 

\begin{cdrfigure}[LBNF organization]{lbnf-org}{LBNF organization}
  \includegraphics[width=0.8\textwidth]{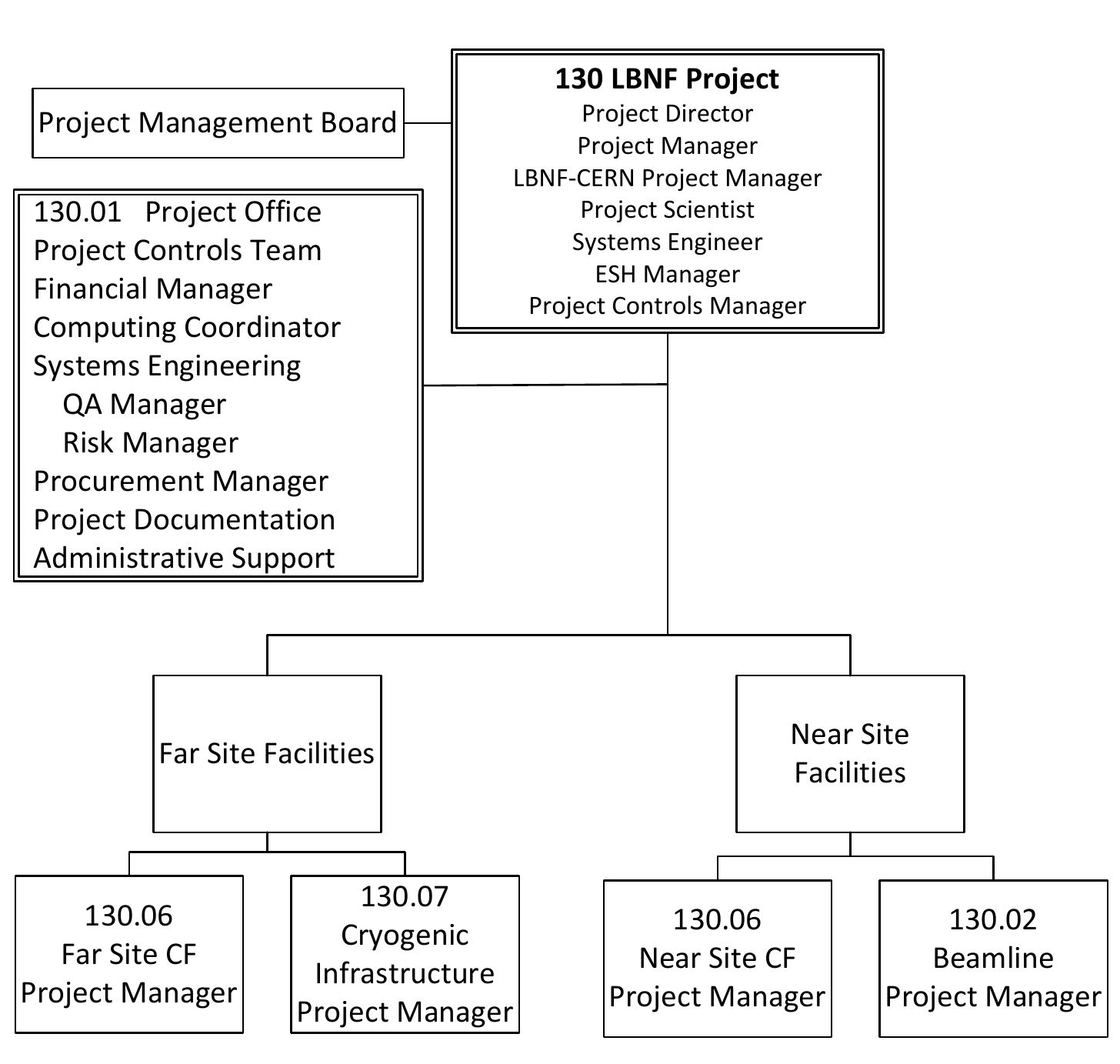}
\end{cdrfigure}



\subsection{SDSTA and SURF}

LBNF plans to construct facilities at SURF to house the DUNE far detector. SURF is owned by the state of South Dakota and managed by the SDSTA. 

Current SURF activities include operations necessary for allowing safe access to the 4850L of the mine, which houses the existing and under-development science experiments. The DOE is presently funding SDSTA ongoing operations through Lawrence Berkeley National Laboratory (LBNL) and its SURF Operations Office through FY16; this is expected to change to funding through Fermilab starting in FY17. 

The LBNF Far Site Facilities Manager is also an employee of SDSTA and is contracted to Fermilab to provide management and coordination of the Far Site Conventional Facilities (CF) and Cryogenics Infrastructure subprojects. LBNF contracts directly with SDSTA for the design of the required CF at SURF, whereas the actual construction of the CF will be directly contracted from Fermilab. Coordination between SDSTA and the LBNF project is necessary to ensure efficient operations at SURF. This will be facilitated via an agreement being developed between SDSTA and Fermilab regarding the LBNF project \fixme{[new reference]} that defines responsibilities and methods for working jointly on LBNF project design and construction. A separate agreement will be written for LBNF operations. 

\subsection{CERN}

The European Organization for Nuclear Research (CERN) is expected to significantly contribute to LBNF with technical components, required to support the deployment of the DUNE detectors and the neutrino beamline. 

\subsection{Coordination within LBNF}

The LBNF project organization is headed by the LBNF Project Director who is also the Fermilab Deputy Director for LBNF and reports directly to the Fermilab Director. 
Within Fermilab's organization, two new divisions are being created to execute the Far Site Facilities and Near Site Facilities subprojects. The heads of these divisions will report to the LBNF Project Manager. 
Any personnel working more than half-time on these subprojects would typically be expected to become a member of one of these divisions, while other contributors will likely be matrixed in part-time roles from other Fermilab Divisions.  The heads of the other Fermilab Divisions work with the L1 and L2 project managers to supply the needed resources on an annual basis.  The management structure described above is currently being transitioned into and will not be fully in place until the Fall of 2015.  

The LBNF WBS defines the scope of the work. All changes to the WBS must be approved by the LBNF Project Manager prior to implementation. At the time of CD-1-Refresh, the LBNF WBS is in transition. Both the current and the post CD-1-R WBS is shown in Figure~\ref{fig:lbnf-wbs} to demonstrate how the scope will map from one WBS to the other. 
SDSTA assigns engineers and others as required to work on specific tasks required for the LBNF project at the SURF site. This is listed in the resource-loaded schedule as contracted work from Fermilab for Far Site CF activities. 
CERN and Fermilab are developing a common cryogenics team to design and produce the Cryogenics Infrastructure subproject deliverables for the far site. CERN provides engineers and other staff as needed to complete their agreed-upon deliverables.  
LBNF has formed several management groups with responsibilities as described below.

\textbf{Project Management Board:} LBNF uses a Project Management Board to provide formal advice to the Project Director on matters of importance to the LBNF project as a whole. Such matters include (but are not limited to) those that:
\begin{itemize}
\item have significant technical, cost, or schedule impact on the project,
\item have impacts on more than one L2 subproject,
\item affect the management systems for the project,
\item have impacts on or result from changes to other projects on which LBNF is dependent, and/or
\item result from external reviews or reviews called by the Project Director
\end{itemize}
The Management Board serves as the
\begin{itemize}
\item LBNF Change Control Board, as described in the Configuration Management Plan\cite{CMP-10760}, and the 
\item Risk Management Board, as described in the Fermilab Risk Management Procedure for Projects~\cite{fnal-risk-mgmt}. 
\end{itemize}

\textbf{Beamline Technical Board:} The role of the LBNF Beamline Technical Board (TB) is to provide recommendations and advice to the Beamline Project Manager on important technical decisions that affect the design and construction of the Beamline. The members of the Technical Board must have knowledge of the project objectives and priorities in order to perform this function. The Beamline Project Manager chairs the Beamline TB. The Beamline Project Engineer is the Scientific Secretary of the Board and co-chairs the Beamline TB as needed. 

\textbf{FSCF Neutrino Cavity Advisory Board:} The Far Site CF (FSCF) project has engaged three international experts in hard rock underground construction to advise it periodically through the design and construction process regarding excavation at SURF. The Board meets at the request of the FSCF-PM, generally on site to discuss specific technical issues. The Board produces a report with its findings and conclusions for project information and action. 

\section{DUNE}

\subsection{DUNE Collaboration Structure}

The DUNE collaboration brings together the members of the international science community
interested in participating in the DUNE experiment.  The collaboration defines the scientific goals of the experiment and subsequently the requirements on the experimental facilities needed to achieve these goals.  The collaboration also provides the scientific effort required for the design and construction of the DUNE detectors, operation of the experiment, and analysis of the collected data. There are four main entities within the DUNE organizational structure:

\begin{itemize}
\item DUNE Collaboration, including the General Assembly of the collaboration and the Institutional Board. 
\item DUNE Management, consisting of the two Co-Spokespersons, the Technical Coordinator, and the Resource Coordinator.  These four along with the chair of the Institutional Board and five additional members of the collaboration form the DUNE Executive Committee.
\item DUNE Project Management, containing the Project Office, headed by the Project Manager, and the managers of the DUNE detector and prototyping groups. 
\item DUNE Science Coordination, incorporating the coordinators of the DUNE detector and prototyping groups, the Physics and Software/Computing Coordinators, as well as the DUNE Technical and Finance Boards.
\end{itemize}
The connections between the different members of these entities are illustrated in Figure~\ref{fig:dune-org}.

\begin{cdrfigure}[DUNE Project and Collaboration organization]{dune-org}{DUNE Project and Collaboration organization}
  \includegraphics[width=0.95\textwidth]{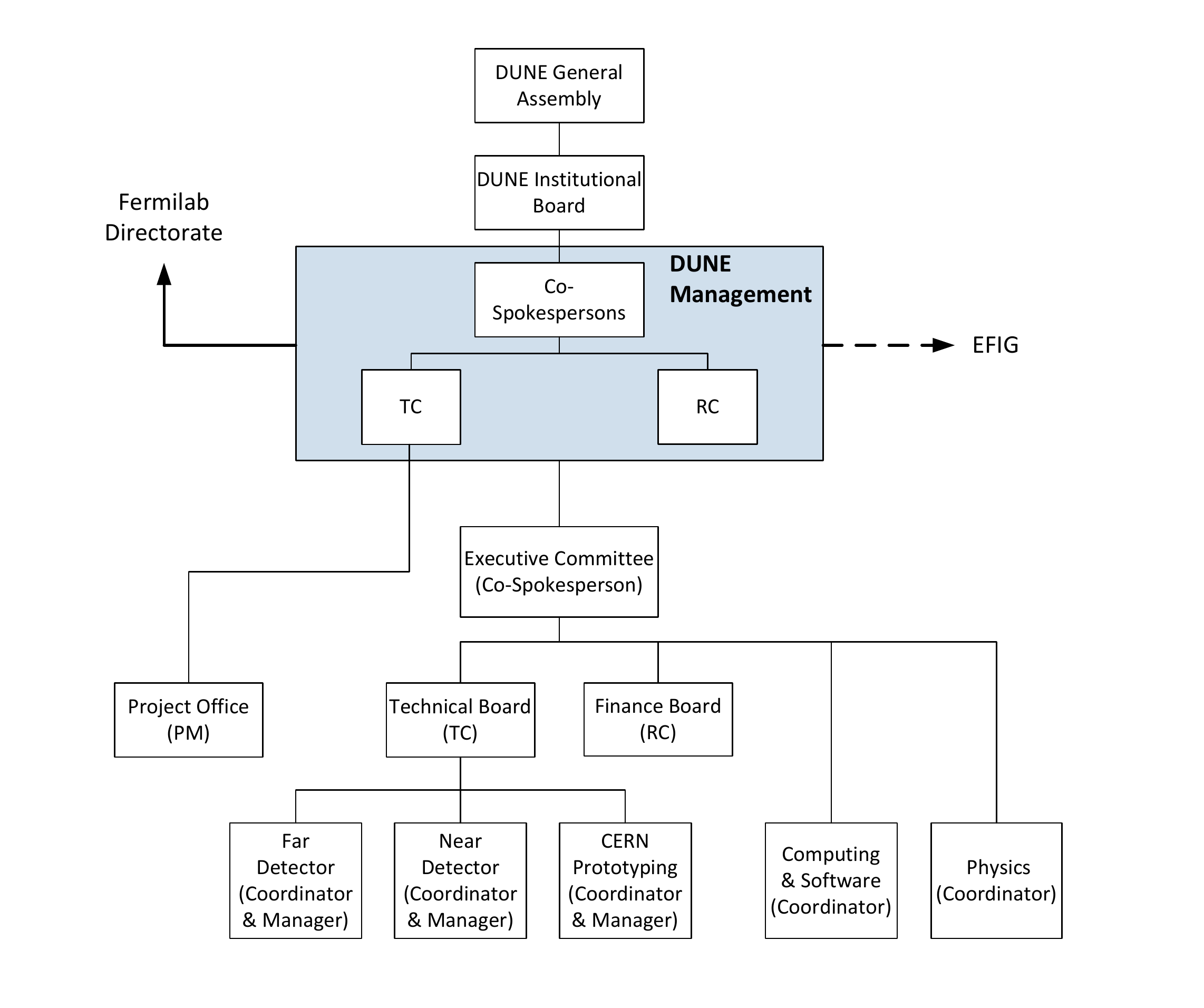}
\end{cdrfigure}

\subsection{DUNE Management Structure}

The main responsibilities of each of the roles are summarized below:
\begin{itemize}
 \item \textbf{DUNE General Assembly} is composed of the full membership of the collaboration.  It is consulted on all major strategic decisions through open plenary sessions at collaboration meetings and is provided regular updates on issues affecting the collaboration at weekly collaboration meetings.  The Collaboration General Assembly elects the Co-Spokespersons through a process defined by the Institutional Board.
\item \textbf{DUNE Institutional Board (IB)} is the representative body of the collaboration institutions. It has responsibility for the governance of the collaboration. The IB has final authority over collaboration membership issues and defines the requirements for inclusion of individuals on the DUNE author list. The IB is also responsible for the process used to select the Co-Spokespersons and the Executive Committee.  The IB chairperson serves on the Executive Committee and runs the Institutional Board meetings.
\item \textbf{DUNE Co-Spokespersons} are elected by the collaboration to serve as its leaders.  They direct collaboration activities on a day-to-day basis and represent the collaboration in interactions with the host laboratory, funding agencies, and the broader scientific community.
\item \textbf{DUNE Executive Committee (EC)} is the primary decision-making body of the collaboration and is chaired by the longest serving Co-Spokesperson.  The membership of the EC consists of the Co-Spokespersons, the Technical Coordinator, the Resource Coordinator, the chair of the IB, and five additional members of the collaboration (three elected IB representatives and two additional members selected by the Co-Spokespersons).  The EC operates as a decision-making body through consensus.  In cases where the EC is unable to reach a consensus, final decision-making authority is assigned to the Co-Spokespersons.  If the Co-Spokespersons are unable to reach their own consensus, the Fermilab Director will step in to resolve the issue.
\item \textbf{Technical Coordinator (TC)} is jointly appointed by the Co-Spokespersons and the Fermilab Director and has reporting responsibilities to both.  In the context of the international DUNE project, the TC serves as the project director 
and is responsible for implementing the scientific and technical strategy of the collaboration.  Currently, the TC also serves as project director for the DOE-funded portion of the DUNE project.  In addition to managing the Project Office, the TC chairs the collaboration Technical Board which coordinates activities associated with the design, construction, installation, and commissioning of the 
detector elements.
\item \textbf{Technical Board (TB)} is chaired by the TC and has a membership that includes the coordinators and managers of the collaboration detector and prototyping groups.  It may also include additional members of the collaboration, nominated by the TC and approved by the EC, who are expected to bring useful knowledge and expertise to its discussions on technical issues.  The TB is the primary 
forum for discussion of issues related to 
detector design, construction, installation and commissioning. 
This body serves as a project change-control board for change requests with schedule and cost impacts that lie below pre-determined thresholds necessitating EC approval.  Change requests that have impacts on interfaces with the LBNF project, 
potential impacts on DUNE science requirements, or that require modifications of formal Memoranda of Understanding (MOU) with one or more 
contributing funding agencies, are discussed within the TB; however these require higher-level approvals, starting with the EC.  The TB is also the primary forum for discussing technological design choices faced by the collaboration.  Based on these discussions, the TB is expected to make a recommendation on the preferred technology choice to the TC, who is then charged with making a final recommendation to the EC. 
\item \textbf{Resource Coordinator (RC)} is jointly appointed by the Co-Spokespersons and the Fermilab Director and has reporting responsibilities to both.  The RC chairs the Collaboration Finance Board and is tasked with preparing the formal MOUs that define the contributions and responsibilities of each institution.  The RC is also responsible for management of the common financial resources of the collaboration (common fund).  Project change requests approved by the EC that involve modification of MOUs with one or more of the participating funding agencies are taken by the RC first to the Collaboration Finance Board for discussion and then, in cases where consensus is obtained, to the Resources Review Board for final approval.
\item \textbf{Finance Board (FB)} is chaired by the RC and has a membership that includes a single representative from each group of collaborating institutions whose financial support for participating in the DUNE experiment originates from a single, independent funding source.  These collaboration representatives are either nominated through their respective group of institutions and approved by the associated funding agency, or directly appointed by the funding agency.  The FB discusses issues related to collaboration resources such as contributions to project common funds and division of project responsibilities among the collaborating institutions.  The FB is also responsible for vetting proposed project change requests prior to their submission to the Resource Research Board for approval.
\item \textbf{DUNE Science Coordinators} include the coordinators of the detector and prototyping groups as well as the coordinators of the DUNE physics and computing/software efforts.  Science coordinators are nominated by the Co-Spokespersons (jointly with the TC in the case of detector and prototyping group coordinators) and approved by the EC.  These coordinators are expected to establish additional collaboration sub-structures within their assigned areas to cover the full scope of collaboration activities within their areas of responsibility.  Detector and prototyping group coordinators report to the EC through the TB, while coordinators of the physics and software/computing efforts report directly to the EC.

\item \textbf{DUNE Project Office (PO)} provides the project management for the design, construction, installation, and commissioning of the DUNE near and far detectors.  Members of the Project Office, including the Project Manager (PM), are appointed by the  TC.  The DUNE project will be run as an international project following DOE guidelines.  The PO will have control over common funds collected from the U.S. and international stakeholders.  Other contributions to the DUNE project are expected to be in the form of deliverables as defined through formal MOUs. The PO will maintain a full schedule for the entire DUNE project and track contributions through detailed subproject milestones.  The entire DUNE project (including international contributions) will follow the DOE critical decision process incorporating a CD-2 approval of its baseline cost and schedule and a CD-3 approval for moving forward with construction.  The current high-level WBS structure of the DUNE project, which will be evolving in the near future to best take advantage of the additional resources available within the new collaboration, is illustrated in Figure~\ref{fig:dune-wbs}.

\begin{cdrfigure}[DUNE Work Breakdown Structure (WBS)]{dune-wbs}{DUNE Work Breakdown Structure (WBS)}
  \includegraphics[width=0.8\textwidth]{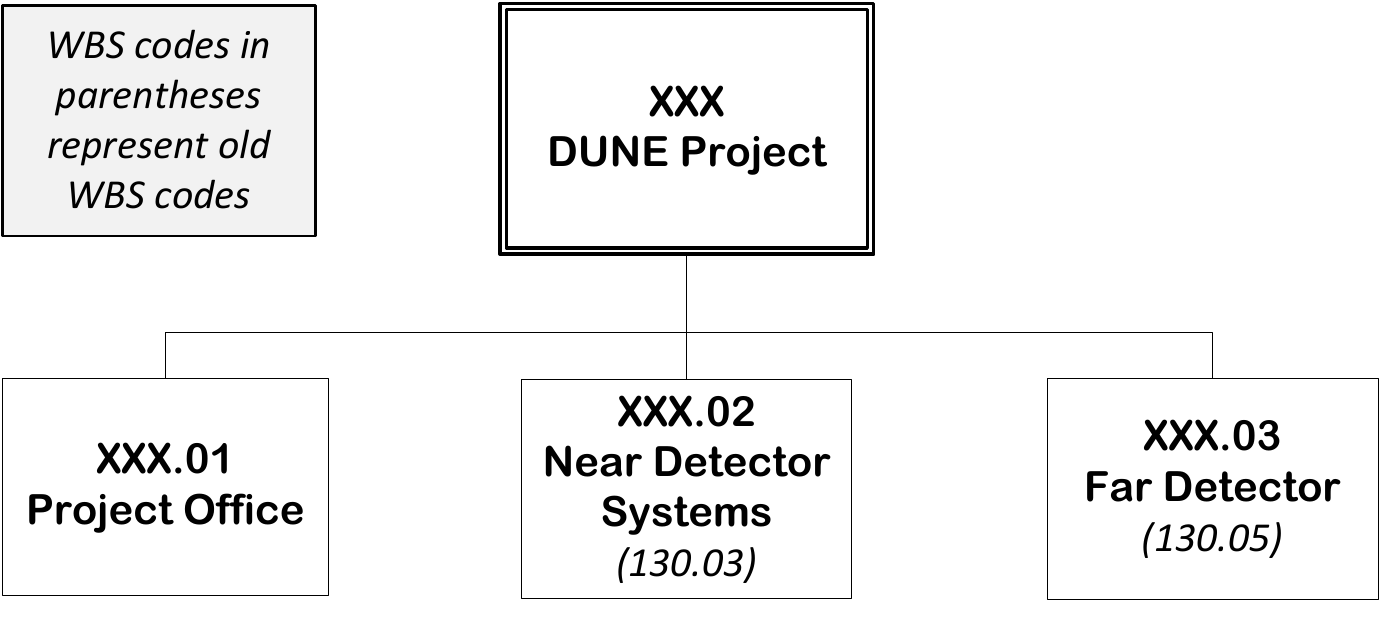}
\end{cdrfigure}

\item \textbf{DUNE Detector and Prototyping Managers} provide the required interface between the DUNE project and the members of the collaboration contributing to these efforts.  These managers sit within the detector and prototyping groups where all matters related to the design, construction, installation, and commissioning of the individual detector elements are discussed.  These managers are tasked with implementing the plans developed within their group and are part of a joint management team which addresses issues associated with project interfaces and coordination of detector and prototyping group efforts.
\end{itemize}

\section{LBNF/DUNE Advisory and Coordinating Structures}
\label{sec:lbnf-dune-interface}

A set of structures is established to provide coordination among the participating funding agencies, oversight of the LBNF and DUNE projects, and coordination and communication between the two projects.  These structures and the relationships among them are shown in Figure~\ref{fig:lbnfdune-org} and are described in this section.

\begin{cdrfigure}[Joint LBNF/DUNE management structure]{lbnfdune-org}{Joint LBNF/DUNE management structure}
  \includegraphics[width=0.8\textwidth]{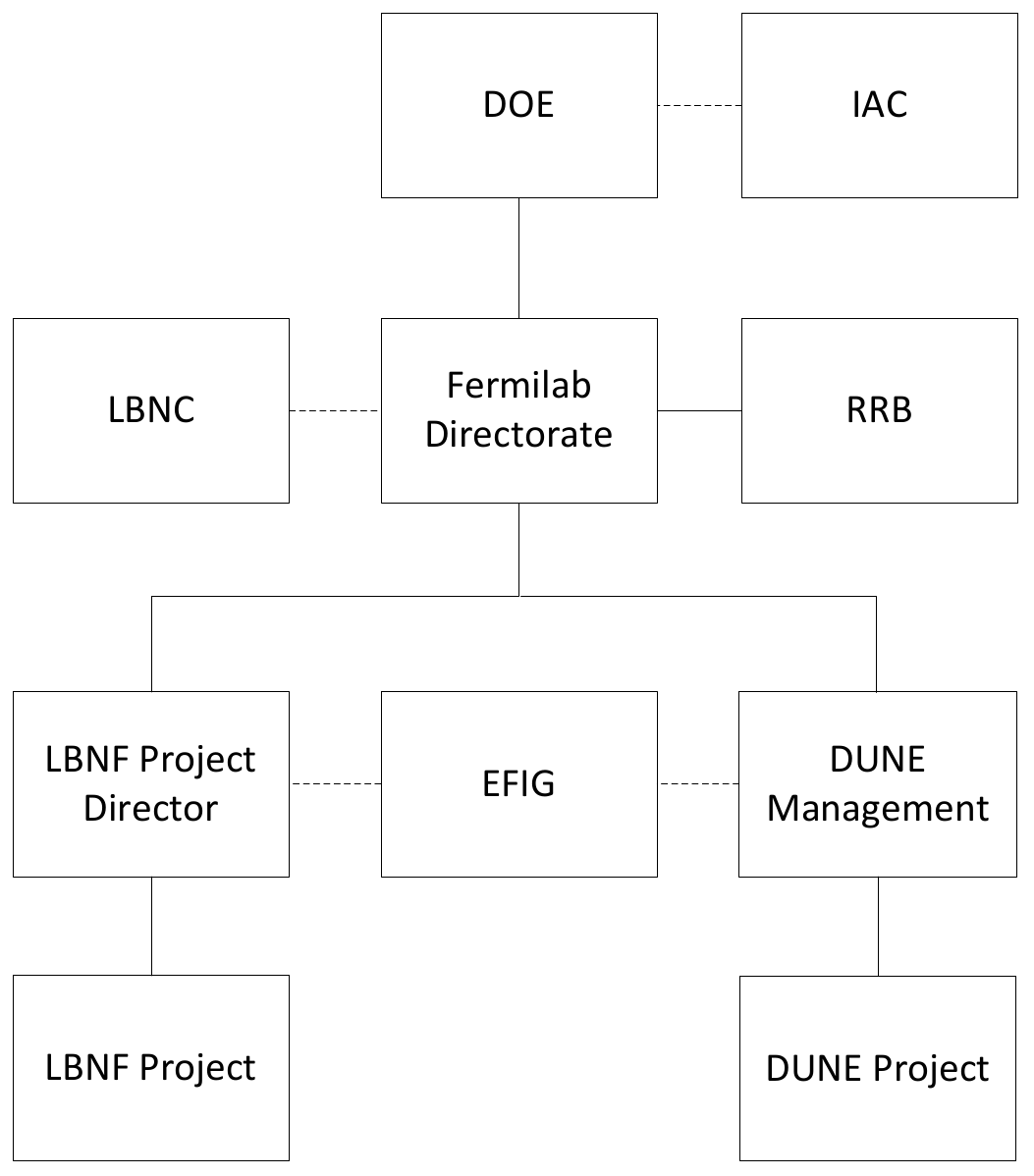}
\end{cdrfigure}

\subsection{International Advisory Council (IAC) }

The International Advisory Council (IAC) is composed of
regional representatives, such as CERN, and representatives of
funding agencies that make major contributions to LBNF infrastructure or to DUNE. The IAC 
acts as the highest-level international advisory body to the U.S.
DOE and the FNAL Directorate and facilitates
high-level global coordination across the entire enterprise (LBNF and DUNE).
The IAC is chaired by the DOE Office of Science Associate Director
for High Energy Physics and includes the FNAL Director in its membership.  
The council meets as needed and provides pertinent advice to 
LBNF and DUNE  
through the Fermilab Director.

Specific responsibilities of the IAC include, but are not limited to,
the following:

\begin{itemize}
\item During the formative stages of LBNF and DUNE
the IAC helps to coordinate the sharing of responsibilities among
the agencies for the construction of LBNF and DUNE.
Individual agency responsibilities for LBNF will be established in
bilateral international agreements with the DOE. Agency contributions to
DUNE will be formalized through separate agreements.

\item The IAC assists in resolving issues, especially those
that cannot be resolved at the Resources Review Boards (RRB) level,
e.g., issues that require substantial redistributions of
responsibilities among the funding agencies.

\item The IAC assists as needed in the coordination,
synthesis and evaluation of input from project reports charged by
individual funding agencies, LBNF and DUNE project management,
and/or the IAC itself, leading to recommendations for action by
the managing bodies.
\end{itemize}

The initial membership, as of May 19, 2015, of the IAC is as follows:
James Siegrist (DOE HEP, Chair),
Sergio Bertolucci (CERN),
Arun Srivastava (DAE),
Carlos Henrique de Brito Cruz (FAPESP),
Fernando Ferroni (INFN),
Fabiola Gianotti (CERN),
Rolf Heuer (CERN),
Stavros Katsanevas (ApPEC),
Frank Linde (ApPEC),
Nigel Lockyer (FNAL),
Reynald Pain (IN2P3/CNRS),
John Womersley (STFC) and
Agnieszka Zalewska (IFJ).

The DUNE Co-Spokespersons and/or other participants within
the Fermilab neutrino program will be invited to sessions of the IAC as
needed. Council membership may increase as additional funding agencies
from certain geographic regions make major contributions to LBNF and DUNE.

\subsection{Resources Review Boards (RRB)}

The Resources Review Boards (RRB) are composed of representatives of all
funding agencies that sponsor LBNF and DUNE, and of the Fermilab
management. The RRB provides focused monitoring and detailed oversight
of each of the projects. The Fermilab Director in coordination
with the DUNE RC defines its membership. A representative from the
Fermilab Directorate chairs the boards and
organize regular meetings to ensure the flow of resources needed
for the smooth progress of the enterprise 
and for its successful completion.  
The managements of the
DUNE collaboration and the LBNF project participates in the RRB meetings
and make regular reports to the RRB on technical, managerial,
financial and administrative matters, as well as status and
progress of the DUNE collaboration.

There are two groups  
within the RRB: RRB-LBNF and RRB-DUNE. Each of
these groups monitors progress and addresses 
 the issues specific to its area 
 while the whole RRB deals with matters
that concern the entire enterprise. 
The RRB will meet
biannually; these meetings 
will start with a plenary
opening session 
and be followed by 
RRB-LBNF and RRB-DUNE sessions. As DUNE progresses toward
experimental operations, RRB-Computing sessions will convene.

DUNE Finance Board members who serve as National Contacts from the 
sponsoring funding agencies will be invited to RRB sessions.

The RRB  employs standing DUNE and LBNF \textit{Scrutiny Groups} as needed
to assist in its responsibilities. The scrutiny groups operate
under the RRB, and provide detailed information on financial and
personnel resources, costing, and other elements under the purview of the RRB.

Roles of the RRB include:

\begin{itemize}
\item assisting the DOE and the FNAL Directorate
with coordinating and developing any required international
agreements between partners,
\item monitoring and overseeing the Common Projects and the
use of the Common Funds,
\item monitoring and overseeing general financial and personnel support,
\item assisting the DOE and the FNAL Directorate
with resolving issues that may require reallocation of responsibilities
among the project's funding agencies,
\item reaching consensus on a maintenance and operation procedure,
and monitoring its function, and
\item  approving the annual construction, and maintenance and operation common fund 
budget of DUNE.
\end{itemize}

\subsection{Fermilab, the Host Laboratory}

As the host laboratory, Fermilab has a direct responsibility for the design,
construction, commissioning and operation of the facilities and
infrastructure (LBNF) that support the science program. 
In this capacity, Fermilab reports
directly to the DOE through the Fermilab Site Office (FSO).
Fermilab also has an important oversight role for the DUNE project
itself as well as an important coordination role in ensuring that
interface issues between the two projects are completely understood.

Fermilab's oversight of the DUNE collaboration and detector
construction project is carried out through
\begin{itemize}
\item regular meetings with the collaboration leadership,
\item approving the selection of collaboration Co-Spokespersons,
\item  providing the Technical and Resource Coordinators,
\item  convening and chairing the Resources Review Boards,
\item  regular scientific reviews by the PAC and LBNC,
\item  Director's Reviews of specific management, technical,
cost and schedule aspects of the detector construction project, and
\item other reviews as needed.
\end{itemize}

\subsection{DUNE Collaboration}	

The collaboration, in consultation with the Fermilab Director,
is responsible for forming the international DUNE project team that is
responsible for designing and constructing the detectors.  
The Technical Coordinator
(TC) and Resource Coordinator (RC) serve as the lead managers
of this international project team and are selected jointly by
the Co-Spokespersons and the Fermilab Director.  Because the international DUNE
project incorporates contributions from a number of different
funding agencies, it 
is responsible for
satisfying individual tracking and reporting requirements associated
with 
the different contributions.

\subsection{Long-Baseline Neutrino Committee (LBNC)}

The Long-Baseline Neutrino Committee (LBNC), composed
of internationally prominent scientists with relevant expertise,
provides external scientific peer review for LBNF and DUNE 
regularly.
The LBNC reviews the scientific, technical and managerial
decisions and preparations for the neutrino program.
It acts in effect 
as an adjunct to the Fermilab Physics Advisory Committee
(PAC), meeting on a more frequent basis than the PAC.
The LBNC may employ DUNE and LBNF Scrutiny Groups for more
detailed reports and evaluations. The LBNC members are appointed by the
Fermilab Director. The current membership of the LBNC is:
David MacFarlane (SLAC, Chair),
Ursula Bassler (IN2P3),
Francesca Di Lodovico (Queen Mary),
Patrick Huber (Virginia Tech),
Mike Lindgren (FNAL),
Naba Mondal (TIFR),
Tsuyoshi Nakaya (Kyoto),
Dave Nygren (UT Arlington),
Stephen Pordes (FNAL),
Kem Robinson (LBNL),
Nigel Smith (SNOLAB) and
Dave Wark (Oxford and STFC).
Among these members, David McFarlane and Dave Wark are also members of the Fermilab PAC.

\subsection{Experiment-Facility Interface Group (EFIG)}

Close and continuous coordination between DUNE and LBNF is
required to ensure the success of the combined enterprise.
An Experiment-Facility Interface Group (EFIG) was established
in January 2015 to oversee and ensure the required coordination
both during the design/construction and operational
phases of the program. This group covers areas including:
\begin{itemize}
\item  interface between the near and far detectors and the
corresponding conventional facilities,
\item interface between the detector systems provided by
DUNE and the technical infrastructure provided by LBNF, and 
\item design and operation of the LBNF neutrino beamline.
\end{itemize}

The EFIG is chaired by two deputy directors of Fermilab.
Its membership includes the LBNF Project Director, Project Manager and Project Scientist, and 
the DUNE Co-Spokespersons, Technical Coordinator, Resource Coordinator and the CERN-LBNF Project Manager.
In consultation with the DUNE and LBNF management, the EFIG Chairs will
extend the membership as needed 
to carry out the coordination
function. In addition, the DOE Federal Project Director for LBNF,
the Fermilab Chief Project Officer, and a designated representative
of the South Dakota Science and Technology Authority (SDSTA) will
serve ex officio. The EFIG Chairs designate a Secretary of the EFIG,
who keeps minutes of the meetings and performs other tasks as
requested by the Chair.

It is the responsibility of the EFIG Chairs to report EFIG proceedings
to the Fermilab Director and other stakeholders. It is the responsibility
of the DUNE Co-Spokespersons to report EFIG proceedings to the rest of
the collaboration. The EFIG meets weekly or as needed.

The current membership of the EFIG is:
Joe Lykken (representing Fermilab Director, Chair),
Nigel Lockyer (acting LBNF Project Director),
Elaine McCluskey (LBNF Project Manager),
Jim Strait (LBNF Project Scientist),
\hyphenation{Andr\'e}
Andr\'e Rubbia (DUNE Co-Spokesperson),
Mark Thomson (DUNE Co-Spokesperson),
Eric James (DUNE Technical Coordinator),
Chang Kee Jung (DUNE Resource Coordinator),
Marzio Nessi (CERN),
David Lissauer (BNL),
Jim Stewart (BNL),
Jeff Dolph (BNL, Secretary),
Mike Lindgren (FNAL Chief Project Officer, ex officio),
Pepin Carolan (DOE, ex officio), and 
Mike Headley (SDSTA, ex officio).

%% file: volume-project/chapter-summary.tex
\chapter{Summary}
\label{ch:project-summary}

LBNF/DUNE will be a world-leading facility for pursuing a cutting-edge program of neutrino physics and astroparticle physics. The 
combination of the intense wide-band neutrino beam, the massive LArTPC far detector and the highly capable near detector will provide 
the opportunity to discover CP violation in the neutrino sector as well as to determine the neutrino mass ordering and provide a 
precision test of the three-flavor oscillation paradigm. The massive, deep-underground far detector will offer unprecedented sensitivity for 
theoretically favored proton decay modes 
and for observation of electron neutrinos from a core-collapse supernova, should one occur in our galaxy during the operation of the experiment.

In addition to summarizing the compelling scientific case for LBNF/DUNE, this document presents an overview of the technical 
designs of the facility and experiment and the strategy for their implementation. 
This strategy delivers the science goals described in the 2014 report of the Particle Physics Project Prioritisation Panel (P5) on a 
competitive timescale. Furthermore, a detailed management plan for the 
organization of LBNF as a U.S.-hosted facility and the DUNE experiment 
as a broad international scientific collaboration has been developed, thus satisfying the goal of internationalizing the project as highlighted in the 
P5 report.

%% file: common/final.tex

\cleardoublepage
\renewcommand{\bibname}{References}
\bibliographystyle{ieeetr}
\bibliography{common/citedb}

%% file: volume-project.bbl
\begin{thebibliography}{10}

\bibitem{pip2-2013}
D.~P. {\em et~al.}, ``{Proton Improvement Plan-II}.''
  \url{http://projectx-docdb.fnal.gov/cgi-bin/RetrieveFile?docid=1232&filename=1.2%20MW%20Report_Rev5.pdf&version=3},
  2013.

\bibitem{p5report2014}
{Particle Physics Project Prioritization Panel}, ``{Building for Discovery;
  Strategic Plan for U.S. Particle Physics in the Global Context},'' 2014.
\newblock
  \url{http://science.energy.gov/~/media/hep/hepap/pdf/May%202014/FINAL_P5_Report_Interactive_060214.pdf}.

\bibitem{ESPP-2012}
\url{http://espp2012.ifj.edu.pl/index.php}.

\bibitem{Huber:2010dx}
P.~Huber and J.~Kopp, ``{Two experiments for the price of one? -- The role of
  the second oscillation maximum in long baseline neutrino experiments},'' {\em
  JHEP}, vol.~1103, p.~013, 2011.

\bibitem{kearns_isoups}
E.~Kearns, ``{Future Experiments for Proton Decay. Presentation at ISOUPS
  (International Symposium: Opportunities in Underground Physics for Snowmass),
  Asilomar, May 2013},'' 2013.

\bibitem{lbnecdr}
{LBNE Collaboration}, ``{The LBNE Conceptual Design Report}.'' LBNE DocDB 5235,
  4317, 4724, 4892, 4623, 5017, 2012.

\bibitem{radiate-web}
\url{http://radiate.fnal.gov/}.

\bibitem{PMP-10770}
{LBNF Project Office}, ``{LBNF Project Management Plan},'' tech. rep., FNAL,
  2015.
\newblock LBNF Doc 10770.

\bibitem{CMP-10760}
{LBNF Project Office}, ``{LBNF Configuration Management Plan},'' tech. rep.,
  FNAL, 2015.
\newblock LBNF Doc 10760.

\bibitem{fnal-risk-mgmt}
``{Fermilab Risk Management Procedure for Projects},'' 2015.
\newblock
  \url{http://www.fnal.gov/directorate/OPMO/PolProc/Fermilab-Risk-Management-Procedure-v1-0-Signed.pdf}.

\end{thebibliography}
